\documentclass[a4paper,fleqn,usenatbib]{mnras}

\usepackage[T1]{fontenc}
\usepackage{ae,aecompl}
\usepackage{fixltx2e}

\usepackage{graphicx}	
\usepackage{amsmath}	
\usepackage{amssymb}	
\usepackage{comment}
\usepackage{ulem}
\usepackage{subcaption}%

\usepackage{xcolor}		
\usepackage{mathrsfs} 
\usepackage{breakcites} 
\usepackage{caption} 
\usepackage{multicol}
\usepackage{rotating} 
\usepackage{threeparttable}
\usepackage{float}

\usepackage{dblfloatfix}

\usepackage{letltxmacro} 
\usepackage{xparse}

\usepackage{newtxtext,newtxmath}

\newcommand{\todo}{\ifmmode {\color{red}\text{\Huge{\(\bullet\)}}} \else {\color{red} \Huge$\bullet$}\fi}
\newcommand{\tido}{\ifmmode {\bullet} \else $\bullet$\fi}

\newcommand{\E        }[1]{\ifmmode 10^{#1} \else $10^{#1}$\fi}
\newcommand{\tE        }[1]{\ifmmode \times10^{#1} \else $\times10^{#1}$\fi}
\newcommand{\til}{\ifmmode \sim \else $\sim$\fi}
\renewcommand{\~} {\ifmmode \sim \else $\sim$\fi}

\newcommand{\pc}	{\ifmmode {\rm pc} \else pc\fi}
\newcommand{\ld}	{\ifmmode {\rm l.d.} \else l.d.\fi}
\newcommand{\kms}	{\ifmmode {\rm km\,s}^{-1} \else km\,s$^{-1}$\fi}
\newcommand{\Jykms}	{\ifmmode {\rm Jy\,km\,s}^{-1} \else Jy\,km\,s$^{-1}$\fi}
\newcommand{\cc}	{\ifmmode {\rm cm}^{-3}    \else cm$^{-3}$\fi}
\newcommand{\cmii}	{\ifmmode {\rm cm}^{-2}    \else cm$^{-2}$\fi}
\newcommand{\ergs}	{\ifmmode {\rm erg\,s}^{-1} \else erg s$^{-1}$\fi}
\newcommand{\ergcms}	{\ifmmode {\rm erg\,cm}^{-2}\,{\rm s}^{-1} \else erg\,cm$^{-2}$\,s$^{-1}$\fi}
\newcommand{\ergcmsA}	{\ifmmode {\rm erg\,cm}^{-2}\,{\rm s}^{-1}\,{\rm\AA}^{-1}
\else erg\,cm$^{-2}$\,s$^{-1}$\,\AA$^{-1}$\fi}
\newcommand{  \ergcmsHz  }{\ifmmode{\rm erg\,cm}^{-2}\,{\rm s}^{-1}\,{\rm Hz}^{-1}
                       \else ergs\,cm$^{-2}$\,s$^{-1}$\,Hz$^{-1}$\fi}
\newcommand{\kev}	{\ifmmode {\rm keV} \else keV\fi}

\newcommand{\mic}	{\ifmmode {\rm \mu m} \else $\mu$m\fi}
\newcommand{\vFWHM}	{\ifmmode v_{\mbox{\tiny FWHM}} \else $v_{\mbox{\tiny FWHM}}$\fi}
\newcommand{\vBLR}	{\ifmmode v_{\mbox{\tiny BLR}} \else $v_{\mbox{\tiny BLR}}$\fi}
\newcommand{\sigBLR}	{\ifmmode \sigma_{\mbox{\tiny BLR}} \else $\sigma_{\mbox{\tiny BLR}}$\fi}
\newcommand{\vNLR}	{\ifmmode v_{\mbox{\tiny NLR}} \else $v_{\mbox{\tiny NLR}}$\fi}
\newcommand{\tauBLR}	{\ifmmode \tau_{\mbox{\tiny BLR}} \else $\tau_{\mbox{\tiny BLR}}$\fi}

\newcommand{\Hubble}	{\ifmmode {\rm km\,s}^{-1}\,{\rm Mpc}^{-1} \else km\,s$^{-1}$\,Mpc$^{-1}$\fi}
\newcommand{\NDunit}	{\ifmmode {\rm Mpc}^{-3} \else Mpc$^{-3}$\fi}
\newcommand{\LFunit}	{\ifmmode {\rm Mpc}^{-3}\,{\rm mag}^{-1} \else Mpc$^{-3}$\,mag$^{-1}$\fi}
\newcommand{\MFunit}	{\ifmmode {\rm Mpc}^{-3}\,{\rm dex}^{-1} \else Mpc$^{-3}$\,dex$^{-1}$\fi}

\newcommand{\Msun}{\ifmmode M_{\odot} \else $M_{\odot}$\fi}
\newcommand{\Lsun}{\ifmmode L_{\odot} \else $L_{\odot}$\fi}
\newcommand{\Zsun}{\ifmmode Z_{\odot} \else $Z_{\odot}$\fi}
\newcommand{\mpyr}{\ifmmode \Msun\,{\rm yr}^{-1} \else $\Msun\,{\rm yr}^{-1}$\fi}

\newcommand{\qnote}{\ifmmode q_{0} \else $q_{0}$\fi}
\newcommand{\Hnote}{\ifmmode H_{0} \else $H_{0}$\fi}
\newcommand{\hnote}{\ifmmode h_{0} \else $h_{0}$\fi}
\newcommand{\anote}{\ifmmode a_{0} \else $a_{0}$\fi}

\def\gsim{\;\rlap{\lower 2.5pt \hbox{$\sim$}}\raise 1.5pt\hbox{$>$}\;}
\def\lsim{\;\rlap{\lower 2.5pt \hbox{$\sim$}}\raise 1.5pt\hbox{$<$}\;}

\newcommand{  \Halpha   }{\ifmmode {\rm H}\alpha \else H$\alpha$\fi}

\newcommand{  \ha       }{\Halpha}
\newcommand{  \Hbeta    }{\ifmmode {\rm H}\beta \else H$\beta$\fi}

\newcommand{  \hb       }{\Hbeta}
\newcommand{  \Hgamma   }{\ifmmode {\rm H}\gamma \else H$\gamma$\fi}
\newcommand{  \Hdelta   }{\ifmmode {\rm H}\delta \else H$\delta$\fi}
\newcommand{  \Lya      }{\ifmmode {\rm Ly}\alpha \else Ly$\alpha$\fi}
\newcommand{  \Lyb      }{\ifmmode {\rm Ly}\beta \else Ly$\beta$\fi}
\newcommand{  \Pa       }{\ifmmode {\rm P}\alpha \else P$\alpha$\fi}
\newcommand{  \Pb       }{\ifmmode {\rm P}\beta \else P$\beta$\fi}
\newcommand{  \Bra      }{\ifmmode {\rm Br}\alpha \else Br$\alpha$\fi}
\newcommand{  \Brg      }{\ifmmode {\rm Br}\gamma \else Br$\gamma$\fi}
\newcommand{  \hi       }{\ifmmode {\rm H}\,\textsc{i} \else H\,\textsc{i}\fi}
\newcommand{  \HI       }{\ifmmode {\rm H}\,\textsc{i} \else H\,\textsc{i}\fi}
\newcommand{  \hii      }{\ifmmode {\rm H}\,\textsc{ii} \else H\,\textsc{ii}\fi}
\newcommand{  \hei      }{\ifmmode {\rm He}\,\textsc{i} \else He\,\textsc{i}\fi}
\newcommand{  \heii     }{\ifmmode {\rm He}\,\textsc{ii} \else He\,\textsc{ii}\fi}
\newcommand{  \HeIIuv   }{\ifmmode {\rm He}\,\textsc{ii}\,\lambda1640 \else He\,\textsc{ii}\,$\lambda1640$\fi}
\newcommand{  \HeIIop   }{\ifmmode {\rm He}\,\textsc{ii}\,\lambda4686 \else He\,\textsc{ii}\,$\lambda4686$\fi}
\newcommand{  \CII	}{\ifmmode \left[{\rm C}\,\textsc{ii}\right]\,\lambda157.74\,\mu{\rm m} \else [C\,{\sc ii}]\ $\lambda157.74\,\mu{\rm m}$\fi}
\newcommand{  \cii	}{\ifmmode \left[{\rm C}\,\textsc{ii}\right] \else [C\,{\sc ii}]\fi}

\newcommand{  \ciii     }{\ifmmode {\rm C}\,\textsc{iii}\right] \else C\,\textsc{iii}]\fi}
\newcommand{  \CIII     }{\ifmmode {\rm C}\,\textsc{iii}\right]\,\lambda1909 \else C\,\textsc{iii}]\,$\lambda1909$\fi}
\newcommand{  \civ      }{\ifmmode {\rm C}\,\textsc{iv}  \else C\,\textsc{iv}\fi}
\newcommand{  \CIV      }{\ifmmode {\rm C}\,\textsc{iv}\,\lambda1549 \else C\,\textsc{iv}\,$\lambda1549$\fi}
\newcommand{  \nii      }{\ifmmode {\rm N}\,\textsc{ii}  \else N\,\textsc{ii}\fi}
\newcommand{  \niii     }{\ifmmode {\rm N}\,\textsc{iii} \else N\,\textsc{iii}\fi}
\newcommand{  \niv      }{\ifmmode {\rm N}\,\textsc{iv}  \else N\,\textsc{iv}\fi}
\newcommand{  \NIVuv    }{\ifmmode {\rm N}\,\textsc{iv}\,\lambda1486 \else N\,\textsc{iv}\,$\lambda1486$\fi}
\newcommand{  \nv       }{\ifmmode {\rm N}\,\textsc{v}   \else N\,\textsc{v}\fi}
\newcommand{\oi}{\ifmmode \left[{\rm O}\,\textsc{i}\right] \else [O\,{\sc i}]\fi}
\newcommand{\OI}{\ifmmode \left[{\rm O}\,\textsc{i}\right]\,\lambda6300 \else [O\,{\sc i}]$\,\lambda6300$\fi}
\newcommand{\oii}{\ifmmode \left[{\rm O}\,\textsc{ii}\right] \else [O\,{\sc ii}]\fi}
\newcommand{\OII}{\ifmmode \left[{\rm O}\,\textsc{ii}\right]\,\lambda3727 \else [O\,{\sc ii}]\,$\lambda3727$\fi}
\newcommand{\oiii}{\ifmmode \left[{\rm O}\,\textsc{iii}\right] \else [O\,{\sc iii}]\fi}
\newcommand{\OIII}{\ifmmode \left[{\rm O}\,\textsc{iii}\right]\,\lambda5007 \else [O\,{\sc iii}]\,$\lambda5007$\fi}
\newcommand{  \OIIIuv   }{\ifmmode {\rm O}\,\textsc{iii}\,\lambda1663 \else O\,\textsc{iii}\,$\lambda1663$\fi}
\newcommand{  \oiv      }{\ifmmode {\rm O}\,\textsc{iv}  \else O\,\textsc{iv}\fi}
\newcommand{  \OIVuv    }{\ifmmode {\rm O}\,\textsc{iv}\,\lambda1402  \else O\,\textsc{iv}\,$\lambda1402$\fi}
\newcommand{  \OIVIR    }{\ifmmode {\rm O}\,\textsc{iv}\,25.9\,\mu {\rm m} \else O\,\textsc{iv}\,$25.9\,\mu$m\fi}
\newcommand{  \ovi      }{\ifmmode {\rm O}\,\textsc{vi}   \else O\,\textsc{vi}\fi}
\newcommand{  \Ovi      }{\ifmmode {\rm O}\,\textsc{vi}\,\lambda1035 \else O\,\textsc{vi}\,$\lambda1035$\fi}
\newcommand{  \nei      }{\ifmmode {\rm Ne}\,\textsc{i}   \else Ne\,\textsc{i}\fi}
\newcommand{  \neii     }{\ifmmode {\rm Ne}\,\textsc{ii}  \else Ne\,\textsc{ii}\fi}
\newcommand{  \NeiiIR   }{\ifmmode {\rm Ne}\,\textsc{ii}\,12.8\,\mu {\rm m} \else Ne\,\textsc{ii}\,$12.8\,\mu$m\fi}
\newcommand{  \neiii    }{\ifmmode {\rm Ne}\,\textsc{iii} \else Ne\,\textsc{iii}\fi}
\newcommand{  \neiv     }{\ifmmode {\rm Ne}\,\textsc{iv}  \else Ne\,\textsc{iv}\fi}
\newcommand{  \nev      }{\ifmmode {\rm Ne}\,\textsc{v}   \else Ne\,\textsc{v}\fi}
\newcommand{  \NevIR    }{\ifmmode {\rm Ne}\,\textsc{v}\,24.3\,\mu {\rm m} \else Ne\,\textsc{v}\,$24.3\,\mu$m\fi}
\newcommand{  \nevi     }{\ifmmode {\rm Ne}\,\textsc{vi}  \else Ne\,\textsc{vi}\fi}
\newcommand{  \mgi      }{\ifmmode {\rm Mg}\,\textsc{i} \else Mg\,\textsc{i}\fi}
\newcommand{  \mgii     }{\ifmmode {\rm Mg}\,\textsc{ii} \else Mg\,\textsc{ii}\fi}
\newcommand{  \MgII     }{\ifmmode {\rm Mg}\,\textsc{ii}\,\lambda2798 \else Mg\,\textsc{ii}\,$\lambda2798$\fi}
\newcommand{  \sii      }{\ifmmode {\rm S}\,\textsc{ii} \else S\,\textsc{ii}\fi}
\newcommand{  \siii     }{\ifmmode {\rm S}\,\textsc{iii} \else S\,\textsc{iii}\fi}
\newcommand{  \siv      }{\ifmmode {\rm S}\,\textsc{iv} \else S\,\textsc{iv}\fi}
\newcommand{  \sili     }{\ifmmode {\rm Si}\,\textsc{i}   \else Si\,\textsc{i}\fi}
\newcommand{  \silii    }{\ifmmode {\rm Si}\,\textsc{ii}  \else Si\,\textsc{ii}\fi}
\newcommand{  \Siliv    }{\ifmmode {\rm Si}\,\textsc{iv}  \else Si\,\textsc{iv}\fi}
\newcommand{  \SilIVuv  }{\ifmmode {\rm Si}\,\textsc{iv}\,\lambda1400  \else Si\,\textsc{iv}\,$\lambda1400$\fi}
\newcommand{  \AlIII   }{\ifmmode {\rm Al}\,\textsc{iii}\,\lambda1857 \else Al\,\textsc{iii}\,$\lambda1857$\fi}
\newcommand{  \Aliii   }{\ifmmode {\rm Al}\,\textsc{iii} \else Al\,\textsc{iii}\fi}
\newcommand{  \caii     }{\ifmmode {\rm Ca}\,\textsc{ii} \else Ca\,\textsc{ii}\fi}
\newcommand{  \feii     }{\ifmmode {\rm Fe}\,\textsc{ii} \else Fe\,\textsc{ii}\fi}
\newcommand{  \feiii    }{\ifmmode {\rm Fe}\,\textsc{iii} \else Fe\,\textsc{iii}\fi}
\newcommand{  \Kalpha   }{\ifmmode {\rm K}\alpha \else K$\alpha$\fi}

\newcommand{ \Lhb   }{\ifmmode L_{\hb} \else $L_{\hb}$\fi}
\newcommand{ \Lha   }{\ifmmode L_{\ha} \else $L_{\ha}$\fi}
\newcommand{ \fwhb  }{\ifmmode {\rm FWHM}\left(\hb\right) \else FWHM(\hb)\fi}
\newcommand{\sighb  }{\ifmmode \sigma\left(\hb\right) \else $\sigma\left(\hb\right)$\fi}
\newcommand{ \ewhb  }{\ifmmode {\rm EW}\left(\hb\right) \else EW(\hb)\fi}
\newcommand{ \fwha  }{\ifmmode {\rm FWHM}\left(\ha\right) \else FWHM(\ha)\fi}
\newcommand{ \ewha  }{\ifmmode {\rm EW}\left(\ha\right) \else EW(\ha)\fi}
\newcommand{ \Lmg   }{\ifmmode L\left(\mgii\right) \else $L\left(\mgii\right)$\fi}
\newcommand{ \fwmg  }{\ifmmode {\rm FWHM}\left(\mgii\right) \else FWHM(\mgii)\fi}
\newcommand{ \Lciv  }{\ifmmode L\left(\civ\right) \else $L\left(\civ\right)$\fi}
\newcommand{ \fwciv }{\ifmmode {\rm FWHM}\left(\civ\right) \else FWHM(\civ)\fi}
\newcommand{ \fwhm  }{\ifmmode {\rm FWHM} \else FWHM\fi} 
\newcommand{ \voff  }{\ifmmode v_{\rm off} \else $v_{\rm off}$\fi} 
\newcommand{ \vmax  }{\ifmmode v_{\rm max} \else $v_{\rm max}$\fi} 

\newcommand{ \mumg  }{\ifmmode \mu\left(\mgii\right) \else $\mu\left(\mgii\right)$\fi}
\newcommand{ \fmg   }{\ifmmode f\left(\mgii\right) \else $f\left(\mgii\right)$\fi}
\newcommand{ \muciv }{\ifmmode \mu\left(\civ\right) \else $\mu\left(\civ\right)$\fi}
\newcommand{ \fciv  }{\ifmmode f\left(\civ\right) \else $f\left(\civ\right)$\fi}

\newcommand{  \auvo     }{\ifmmode \alpha_{\nu,{\rm UVO}} \else $\alpha_{\nu,{\rm UVO}}$\fi}
\newcommand{  \Ledd     }{\ifmmode L_{\rm Edd} \else $L_{\rm Edd}$\fi}
\newcommand{  \lamLlam  }{\ifmmode \lambda L_{\lambda} \else $\lambda L_{\lambda}$\fi}
\newcommand{  \lLl      }{\ifmmode \lambda L_{\lambda} \else $\lambda L_{\lambda}$\fi}
\newcommand{  \nuLnu    }{\ifmmode \nu L_{\nu} \else $\nu L_{\nu}$\fi}
\newcommand{  \nLn      }{\ifmmode \nu L_{\nu} \else $\nu L_{\nu}$\fi}
\newcommand{  \Luv      }{\ifmmode L_{1450} \else $L_{1450}$\fi}
\newcommand{  \Lop      }{\ifmmode L_{5100} \else $L_{5100}$\fi}
\newcommand{  \lLop     }{\ifmmode \log\left(\Lop/\ergs\right) \else $\log\left(\Lop/\ergs\right)$\fi}
\newcommand{  \Lthree   }{\ifmmode L_{3000} \else $L_{3000}$\fi}
\newcommand{  \lLthree  }{\ifmmode \log\left(\Lthree/\ergs\right) \else $\log\left(\Lthree/\ergs\right)$\fi}
\newcommand{  \Lsix      }{\ifmmode L_{6200} \else $L_{6200}$\fi}
\newcommand{  \lLisx     }{\ifmmode \log\left(\Lop/\ergs\right) \else $\log\left(\Lop/\ergs\right)$\fi}
\newcommand{  \Lxray    }{\ifmmode L_{\rm X} \else $L_{\rm X}$\fi}
\newcommand{  \Lhard    }{\ifmmode L_{\rm 2-10} \else $L_{\rm 2-10}$\fi}
\newcommand{  \Lsoft    }{\ifmmode L_{\rm 0.5-2} \else $L_{\rm 0.5-2}$\fi}

\newcommand{\Fthree}{\ifmmode F_{3000} \else $F_{3000}$\fi}
\newcommand{\fuv}{\ifmmode f_{\lambda}\left(1450{\rm \AA}\right) \else $f_{\lambda}\left(1450 {\rm \AA}\right)$\fi}
\newcommand{\fthree}{\ifmmode f_{\lambda}\left(3000{\rm \AA}\right) \else $f_{\lambda}\left(3000{\rm \AA}\right)$\fi}
\newcommand{\fH}{\ifmmode f_{\lambda}\left(1.65\micron\right) \else
$f_{\lambda}\left(1.65\micron\right)$\fi}

\newcommand{\fbol}{\ifmmode f_{\rm bol} \else $f_{\rm bol}$\fi}
\newcommand{\fbolwv}{\ifmmode f_{\rm bol}\left(\lambda\right) \else $f_{\rm bol}\left(\lambda\right)$\fi}
\newcommand{\fbolopt}{\ifmmode f_{\rm bol}\left(5100{\rm \AA}\right) \else $f_{\rm bol}\left(5100{\rm \AA}\right)$\fi}
\newcommand{\fbolthree}{\ifmmode f_{\rm bol}\left(3000{\rm \AA}\right) \else $f_{\rm bol}\left(3000{\rm \AA}\right)$\fi}
\newcommand{\fboluv}{\ifmmode f_{\rm bol}\left(1450{\rm \AA}\right) \else $f_{\rm bol}\left(1450{\rm \AA}\right)$\fi}

\newcommand{\fobs}{\ifmmode f_{\rm obs} \else $f_{\rm obs}$\fi}

\newcommand{  \mbh      }{\ifmmode M_{\rm BH} \else $M_{\rm BH}$\fi}
\newcommand{  \lmbh     }{\ifmmode \log\left(\mbh/\Msun\right) \else $\log\left(\mbh/\Msun\right)$\fi} 
\newcommand{  \lledd    }{\ifmmode L/L_{\rm Edd} \else $L/L_{\rm Edd}$\fi}
\newcommand{  \mmedd    }{\ifmmode \dot{m}/\dot{m}_{\rm \,Edd} \else $\dot{m}/\dot{m}_{\rm \,Edd}$\fi}
\newcommand{  \Lbol     }{\ifmmode L_{\rm bol} \else $L_{\rm bol}$\fi}
\newcommand{  \lbol     }{\ifmmode L_{\rm bol} \else $L_{\rm bol}$\fi}
\newcommand{  \lLbol    }{\ifmmode \log\left(\Lbol/\ergs\right) \else $\log\left(\Lbol/\ergs\right)$\fi} 
\newcommand{  \Lagn     }{\ifmmode L_{\rm AGN} \else $L_{\rm AGN}$\fi}
\newcommand{  \lagn     }{\ifmmode L_{\rm AGN} \else $L_{\rm AGN}$\fi}

\newcommand{  \tgrow     }{\ifmmode t_{\rm growth} \else $t_{\rm growth}$\fi}
\newcommand{  \tUni      }{\ifmmode t_{\rm Universe} \else $t_{\rm Universe}$\fi}

\newcommand{  \Mindot	}{\ifmmode \dot{M}_{\rm infall} \else $\dot{M}_{\rm infall}$\fi}
\newcommand{  \Mbhdot	}{\ifmmode \dot{M}_{\rm BH} \else $\dot{M}_{\rm BH}$\fi}

\newcommand{  \Maddot	}{\ifmmode \dot{M}_{\rm AD} \else $\dot{M}_{\rm AD}$\fi}

\newcommand{  \Mdot	}{\ifmmode \dot{M} \else $\dot{M}$\fi}

\newcommand{  \as	}{\ifmmode a_{\rm *} \else $a_{\rm *}$\fi}
\newcommand{  \avec	}{\ifmmode \vec{a}_{\rm *} \else $\vec{a}_{\rm *}$\fi}
\newcommand{  \re	}{\ifmmode \eta      	 \else $\eta$\fi}
\newcommand{  \RISCO	}{\ifmmode R_{\rm ISCO}  \else $R_{\rm ISCO}$\fi}
\newcommand{  \rg	}{\ifmmode r_{\rm g}  \else $r_{\rm g}$\fi}
\newcommand{  \rS	}{\ifmmode r_{\rm S}  \else $r_{\rm S}$\fi}

\newcommand{  \mseed    }{\ifmmode M_{\rm seed} \else $M_{\rm seed}$\fi}
\newcommand{  \mbul     }{\ifmmode M_{\rm Bulge} \else $M_{\rm Bulge}$\fi} 
\newcommand{  \mstar    }{\ifmmode M_{*} \else $M_{*}$\fi} 
\newcommand{  \mgal     }{\ifmmode M_{*} \else $M_{*}$\fi} 
\newcommand{  \mhost    }{\ifmmode M_{\rm Host} \else $M_{\rm Host}$\fi}
\newcommand{  \mmsmall  }{\ifmmode M_{\rm BH}/M_{*} \else $M_{\rm BH}/M_{*}$\fi}
\newcommand{  \mmlarge  }{\ifmmode M_{*}/M_{\rm BH} \else $M_{*}/M_{\rm BH}$\fi}

\newcommand{  \mmwp     }{\ifmmode \left(M_{*}/M_{\rm BH}\right) \else $\left(M_{*}/M_{\rm BH}\right)$\fi}
\newcommand{  \ml       }{\ifmmode M_{*}/L_{*} \else $M_{*}/L_{*}$\fi}
\newcommand{  \mlwp     }{\ifmmode \left(M_{*}/L\right) \else $\left(M_{*}/L\right)$\fi}
\newcommand{  \mlk      }{\ifmmode \left(M_{*}/L_{K}\right) \else $\left(M_{*}/L_{K}\right)$\fi}
\newcommand{  \sigs     }{\ifmmode \sigma_{*} \else $\sigma_{*}$\fi}
\newcommand{  \Reff     }{\ifmmode R_{\rm e} \else $R_{\rm e}$\fi}
\newcommand{  \nser     }{\ifmmode n_{\rm s} \else $n_{\rm s}$\fi}
\newcommand{  \LFIR     }{\ifmmode L_{\rm FIR} \else $L_{\rm FIR}$\fi}
\newcommand{  \Lfir     }{\ifmmode L_{\rm FIR} \else $L_{\rm FIR}$\fi}

\newcommand{  \mdyn     }{\ifmmode M_{\rm dyn} \else $M_{\rm dyn}$\fi} 
\newcommand{  \mgas     }{\ifmmode M_{\rm gas} \else $M_{\rm gas}$\fi} 
\newcommand{  \mh       }{\ifmmode M_{\rm h} \else $M_{\rm h}$\fi}
\newcommand{  \mhalo    }{\ifmmode M_{\rm halo} \else $M_{\rm halo}$\fi}

\newcommand{  \Lcii     }{\ifmmode L_{\cii} \else $L_{\cii}$\fi}

\newcommand{\bj}{\ifmmode b_{\rm J} \else $b_{\rm J}$\fi}

\newcommand{\iab}{\ifmmode i_{\rm AB} \else $i_{\rm AB}$\fi}

\newcommand{\jab}{\ifmmode J_{\rm AB} \else $J_{\rm AB}$\fi}
\newcommand{\hab}{\ifmmode H_{\rm AB} \else $H_{\rm AB}$\fi}
\newcommand{\kab}{\ifmmode K_{\rm AB} \else $K_{\rm AB}$\fi}

\newcommand{\jveg}{\ifmmode J_{\rm Vega} \else $J_{\rm Vega}$\fi}
\newcommand{\hveg}{\ifmmode H_{\rm Vega} \else $H_{\rm Vega}$\fi}
\newcommand{\kveg}{\ifmmode K_{\rm Vega} \else $K_{\rm Vega}$\fi}

\newcommand{  \Chisq    }{\ifmmode \chi^{2} \else $\chi^{2}$}
\newcommand{  \nelec    }{\ifmmode n_{e} \else $n_{e}$\fi}     
\newcommand{  \nh       }{\ifmmode n_{H} \else $n_{H}$\fi}     
\newcommand{  \Ncol     }{\ifmmode N_{col} \else $N_{col}$\fi} 
\newcommand{  \NH       }{\ifmmode N_{H} \else $N_{H}$\fi}     

\newcommand{\MgasCO}{\ifmmode M_{\rm gas, \, CO} \else $M_{\rm gas, \, CO}$\fi}
\newcommand{\Mgasdust}{\ifmmode M_{\rm gas, \, dust} \else $M_{\rm gas, \, dust}$\fi}
\newcommand{\lMgasCO}{\relax\ifmmode \log \left( M_{\rm gas, \, CO} \right) \else $ \log \left( M_{\rm gas, \, CO}  \right)$\fi}
\newcommand{\lMgasdust}{\ifmmode  \log \left( M_{\rm gas, \, dust} \right) \else $ \log \left( M_{\rm gas, \, dust} \right)$\fi}
\newcommand{\Mdust}{\ifmmode M_{\rm dust} \else $M_{\rm dust}$\fi}
\newcommand{\lMdust}{\ifmmode  \log \left( M_{\rm dust} \right) \else $ \log \left( M_{\rm dust} \right)$\fi}

\newcommand{\Hmol }{\ifmmode {\rm H_2} \else ${\rm H_2}$\fi}

\def\micron{\hbox{$\mu$m}}
\def\ion#1#2{#1$\;${\small\rm\@Roman{#2}}\relax}

\LetLtxMacro\oldcitep\citep 
\RenewDocumentCommand{\citep}{O{} O{} m}{\oldcitep[#1][#2]{#3}}
\NewDocumentCommand{\citex}{O{} O{} m}{\oldcitep{#3}}

\LetLtxMacro\oldcitet\citet 
\RenewDocumentCommand{\citet}{O{} O{} m}{\oldcitet[#1][#2]{#3}}

\newcommand\blfootnote[1]{%
  \begingroup
  \renewcommand\thefootnote{}\footnote{#1}%
  \addtocounter{footnote}{-1}%
  \endgroup
}

\title[3D intrinsic shapes]{3D intrinsic shapes of quiescent galaxies in observations and simulations}

\author[Zhang et al.]{
Junkai Zhang$^{1,{\color{blue} \star}}$,
Stijn Wuyts$^{1}$,
Callum Witten$^{2}$,
Charlotte R. Avery$^{1}$,
Lei Hao$^{3}$, \newauthor
Raman Sharma$^{1}$,
Juntai Shen$^{4,5,3}$,
Jun Toshikawa$^{1}$,
Carolin Villforth$^{1}$
\\
$^{1}$ Department of Physics, University of Bath, Claverton Down, Bath, BA2 7AY, UK\\
$^{2}$ Institute of Astronomy, University of Cambridge, Madingley Road, Cambridge CB3 0HA, UK\\
$^{3}$ Shanghai Astronomical Observatory, 80 Nandan Road, 200030 Shanghai, China\\
$^{4}$ Department of Astronomy, School of Physics and Astronomy, Shanghai Jiao Tong University, 800 Dongchuan Road, Shanghai 200240, China\\
$^{5}$ Key Laboratory for Particle Astrophysics and Cosmology (MOE) / Shanghai Key Laboratory for Particle Physics and Cosmology, Shanghai 200240, China
}

\date{Accepted XXX. Received YYY; in original form ZZZ}

\pubyear{2021}

\begin{document}
\label{firstpage}
\pagerange{\pageref{firstpage}--\pageref{lastpage}}
\maketitle
\raggedbottom  

\begin{abstract}
We study the intrinsic 3D shapes of quiescent galaxies over the last half of cosmic history based on their axial ratio distribution.  To this end, we construct a sample of unprecedented size, exploiting multi-wavelength $u$-to-$K_s$ photometry from the deep wide area surveys KiDS+VIKING paired with high-quality $i$-band imaging from HSC-SSP.  Dependencies of the shapes on mass, redshift, photometric bulge prominence and environment are considered.  For comparison, the intrinsic shapes of quenched galaxies in the IllustrisTNG simulations are analyzed and contrasted to their formation history.  We find that over the full $0<z<0.9$ range, and in both simulations and observations, spheroidal 3D shapes become more abundant at $M_* > 10^{11}\ M_{\odot}$, with the effect being most pronounced at lower redshifts.  In TNG, the most massive galaxies feature the highest ex-situ stellar mass fractions, pointing to violent relaxation via mergers as the mechanism responsible for their 3D shape transformation.  Larger differences between observed and simulated shapes are found at low to intermediate masses.  At any mass, the most spheroidal quiescent galaxies in TNG feature the highest bulge mass fractions, and conversely observed quiescent galaxies with the highest bulge-to-total ratios are found to be intrinsically the roundest.  Finally, we detect an environmental influence on galaxy shape, at least at the highest masses, such that at fixed mass and redshift quiescent galaxies tend to be rounder in denser environments.\\

\end{abstract}


\section{Introduction}
\label{sec:Intro}

\blfootnote{ {\color{blue} $\star$} {E-mail: jz2192@bath.ac.uk}}

The structure of galaxies encodes an imprint of their formation histories.  For most of its century-old history \citep{Hubble1926}, the study of galaxy structure has focussed on morphologies inferred from projected two-dimensional surface brightness distributions on the sky.  These have been found to relate closely to the galaxies' bimodal colour distribution, with star-forming galaxies (SFGs) being mostly characterized by exponential disks, and bulge-dominated surface brightness distributions being more common among quiescent galaxies (QGs) \citep{Strateva2001,Kauffmann2003}.  This is a trend that persists over at least 80\% of cosmic history \citep[e.g.,]{Wuyts2011}.

Mechanisms that are capable of inducing morphological transformation include mergers between galaxies, where the outcome in terms of remnant size and the degree to which disks are destroyed depends significantly on the mass ratio, gas fractions and morphologies of the progenitors, as well as their orbital parameters \citep[e.g.,][]{Barnes1992, Hernquist1992, Naab2003, Naab2006, Cox2006}.  As a rule of thumb, dry and modest mass-ratio (i.e., major) mergers are the most effective at destroying disk components.  At high redshift, violent disk instabilities of marginally stable gas-rich disks have also been proposed as a path to bulge formation \citep{Dekel2009}, with the compact nature of the remnants left by such an early in-situ process demanding additional subsequent evolution to explain the structural scaling relations of today's early-type galaxy population \citep[e.g.,][]{Naab2014}.  Finally, an array of physical processes specific to higher density environments can contribute to morphological transformations, and influence the star-forming versus quiescent status of galaxies, giving rise to  morphology - density and star formation - density relations \citep{Dressler1980, Balogh1998, Gomez2003, Kauffmann2004}.  Among these are ram-pressure stripping by intra-cluster gas, providing a strong `wind' to overcome the gravitational potential of the galaxy and remove the gas within \citep{Gunn1972}; interactions between galaxies such as mergers or galaxy harassment via high-speed impulsive encounters, which can lead to asymmetry, warps, bars and tidal tails \citep{Moore1996}; `strangulation' cutting off the supply of cold gas \citep{Larson1980}; and shape distortions by tidal forces under the gravitation of the cluster and its dark halo  \citep{Byrd1990}. 

Two possible approaches allow going beyond the two-dimensional characterization of galaxy structure.  One avenue adds kinematic information as a third dimension, a technique that has been applied to many thousands of star-forming and quiescent galaxies in the nearby Universe \citep[e.g.,]{Croom2012,Bundy2015,Sanchez2020}, and to samples of over a thousand (predominantly star-forming) galaxies at higher redshifts \citep[see, e.g.,]{ForsterSchreiber2020}.  Among its many successes, such studies revealed the presence of slow- and fast-rotators among nearby quiescent galaxies, even among those sharing the same Hubble type \citep[see, e.g.,][and references therein]{Cappellari2016}.

An alternative approach relies on the selection of a population of galaxies that are deemed to form an ensemble, sharing similar intrinsic properties but being observed from random viewing angles.  Approximating the intrinsic 3D shapes as ellipsoids, their observed projected axial ratio distribution can then be inverted to reveal insights on the (mix of) intrinsic shapes.  Applications to early-type galaxies have been presented by, e.g., \citet{Ryden1992} and \citet{Tremblay1996} for the nearby Universe, and have been pushed to higher redshifts by, e.g., \citet{Chang2013a,Chang2013b}.  In an analysis of star-forming galaxies, \citet{Zhang2019} further emphasized that the robustness of the inversion process is enhanced by considering not just the one-dimensional projected minor-to-major axial ratio $q \equiv b/a$ distribution, but the distribution of galaxies in the projected axial ratio -- projected semi-major axis ($q$ -- $\log(a)$) plane.   Irrespective of its implementation, by its very nature the technique relies on large galaxy samples and thus wide-area surveys.  This is yet more the case if one selects the ensembles by binning according to multiple galaxy properties (e.g., star-forming/quiescent; redshift; stellar mass; bulge-to-total ratio; environment), to evaluate the dependence of intrinsic shape on these parameters and to better ensure that the ensemble can be treated as a homogeneous set of objects viewed from different viewing angles.

Here, we combine three overlapping wide-area surveys: the Kilo-Degree Survey (KiDS; \citealt{Kuijken2019}), the VISTA Kilo-degree Infrared Galaxy (VIKING; \citealt{Edge2013}) and the Hyper Suprime-Cam Subaru Strategic Program (HSC-SSP; \citealt{Aihara2019}) to characterize the redshifts, internal properties and environments of a large ($N_{\rm total} = 2,731,511$, $N_{\rm QG} = 478,677$) sample of galaxies spanning the redshift range $0 < z < 0.9$.  This paper focuses on the projected axial ratios of specifically the quiescent galaxy population, analyzed as a function of mass and redshift, and with further consideration of their surface brightness profiles (as captured by the photometric bulge-to-total ratio $B/T$) and environment (local overdensity $\delta$ and tidal parameter).

For reference, previous analyses of axial ratio distributions of quiescent (or early-type) galaxies in the nearby Universe have considered samples containing several 10,000s to 300,000 objects \citep{Vincent2005, Padilla2008, van2009}.  Studies covering similar redshifts as considered here analysed $\sim 1,300 - 8,100$ objects \citep{Holden2012, Satoh2019}, whereas samples up to $z = 2.5$ counted $\sim 900 - 2,200$ QGs \citep{Chang2013b, Hill2019}, with numbers of $z > 2.5$ QGs with axial ratio measurements in the study by \citet{Hill2019} being restricted to a few 10s.

In this study, we also derive stellar intrinsic 3D shapes for simulated galaxies, using the IllustrisTNG (\citealt{Nelson2018}; hereafter TNG) cosmological hydrodynamical simulation, to shed light on the relation between shape and formation history as well as environment from a theoretical perspective.

The paper is laid out as follows.  After a brief overview of the observed and simulated datasets in Section\ \ref{data.sec}, we describe the methodology used to extract intrinsic 3D shapes for observed and simulated galaxies in Section\ \ref{methodology.sec}, as well the computation of the overdensity and tidal parameter measures of environment.  Section\ \ref{results.sec} then proceeds to present the results, which are summarized in Section\ \ref{summary.sec}.

Throughout the paper, we adopt a \citet{Chabrier2003} IMF and a flat $\Lambda$CDM cosmology with $\Omega_{\Lambda} = 0.7$, $\Omega_m = 0.3$ and $H_0 =70\ \rm{km}\ \rm{s}^{-1}\ \rm{Mpc}^{-1}$.


\section{Data}
\label{data.sec}

\subsection{KiDS, VIKING and HSC-SSP}
\label{KiDS_VIKING_HSC.sec}

This work makes use of the galaxy information contained within the fourth data release of the Kilo-Degree Survey (KiDS; \citealt{Kuijken2019}), the VISTA Kilo-degree Infrared Galaxy Survey (VIKING; \citealt{Edge2013}) as contained within the KiDS DR4 photometric catalogue, and the second public data release of the Hyper Suprime-Cam Subaru Strategic Program (HSC-SSP; \citealt{Aihara2019}).
The KiDS+VIKING photometry is used to derive redshifts, stellar masses and star formation rates (Section \ref{z_Mstar_type.sec}), and the deeper and higher imaging quality $i$-band observations from HSC-SSP are exploited for their morphological information.

KiDS is an optical $ugri$ wide-field imaging survey covering a total of 1350 square degrees with OmegaCAM on the VLT Survey Telescope.  VIKING is a medium-deep extragalactic survey with VISTA in the $zYJHK_s$ bands, largely overlapping with KiDS.  Together, KiDS+VIKING provide the 9-band optical-to-near-infrared photometric coverage of galaxies' spectral energy distributions that we use to compute their photometric redshifts and estimate their stellar population properties.

HSC-SSP is a wide-field optical imaging survey using the 8.2m Subaru Telescope.  Imaging is taken in five broad-band filters ($grizy$).  For this work, it is specifically the superb image quality of the $i$-band (median seeing of $0\farcs6$) that we leverage.  The depth of the $i$-band imaging in the HSC-SSP Wide layer is 25.9 magnitude (5$\sigma$ limits within 2 arcsec diameter apertures).  As part of the HSC pipeline products, measurements of the bulge-to-total ratio ($B/T$) and the axial ratio ($q \equiv b/a$) are provided, both accounting for point spread function (PSF) convolution.  

The greater depth of HSC $i$-band imaging relative to KiDS implies that the structural measurements employed in our study are based on high signal-to-noise ratio (S/N) data, despite our sample going down to the detection threshold of the KiDS+VIKING catalog.  Specifically, 99.95\% of our sample has $S/N_{i, {\rm HSC}} > 20$, 99.71\% has $S/N_{i, {\rm HSC}} > 30$, and 98.33\% has $S/N_{i, {\rm HSC}} > 50$, sufficiently adequate for determination of the global structural parameters of consideration in this paper.

Within the area of overlap with the CANDELS survey (\citealt{Grogin2011}; a tiny fraction of the overall area covered by HSC, corresponding to CANDELS-COSMOS, CANDELS-UDS and CANDELS-EGS), significantly deeper, higher resolution and longer wavelength imaging is available.  The observed-frame $H$-band imaging offers in principle a more reliable tracer of the stellar distributions, in the presence of potential spatial variations in mass-to-light ($M/L$) ratio.  As a sanity check, we verified that the projected axial ratios of quiescent galaxies derived from the latter dataset by \citet{van2014a} agree well with those provided by HSC-SSP over the dynamic range in mass and redshift considered in this study.  Specifically, we find no evidence for systematic offsets at $S/N_{i, {\rm HSC}} > 30$.  Below $S/N_{i, {\rm HSC}} < 30$, a tendency for HSC to recover rounder projected axial ratios than the reference CANDELS values is observed, at the level of +0.05 for $S/N_{i, {\rm HSC}} \sim 20$ and up to +0.25 for $S/N_{i, {\rm HSC}} < 10$.  Only a tiny proportion of galaxies in our sample fall in this regime, so no additional selection criterion was imposed to weed them out.  Absolute differences between HSC and CANDELS $q$ measurements were quantified as a function of $S/N_{i, {\rm HSC}}$, and used to obtain an empirical measure of the typical error on $q$ when modelling the intrinsic shapes of a sample of galaxies with a particular median $S/N_{i, {\rm HSC}}$.  This resulted in adopted errors on $q$ ranging from 0.02 for subsamples of bright objects (low redshift, high mass) up to 0.06 for subsamples of fainter sources (high redshift, low mass).  We further note that for the latter category of objects the empirically derived errors on $q$ are only modestly (4-20\%) higher than the formal uncertainties taken from the HSC database, but larger factors, up to an order of magnitude, are found between empirical and formal $q$ errors for brighter sources.  In other words, while the formal errors on $q$ become arbitrarily small, the empirical ones do not.

\subsection{Sample selection}
\label{sample.sec}

Our sample is constructed by cross-matching the HSC-SSP and KiDS+VIKING catalogues in their overlapping regions using a 1" search radius. To guarantee a genuine and unambiguous match between entries in the two catalogues, we only use objects with a one-to-one match in both directions (i.e., the nearest counterpart in HSC-SSP within 1" of a KiDS+VIKING source also has that same KiDS+VIKING source as nearest cross-match). The area of overlap between the surveys is composed of a few disjoint regions positioned along a long strip on the sky between $129^{\circ}$ and $227^{\circ}$ in longitude and $-2.3^{\circ}$ to $3.0^{\circ}$ in latitude, making for a total of 257 deg$^2$ with 9-band photometric coverage.

We remove objects that fall in masked sky areas\footnote{These are areas on the sky with unreliable photometry due to the vicinity of bright foreground stars, identified as such in the HSC-SSP public data release.}, and we account for any such masked area later when quantifying counts of neighbouring sources within a cylinder around a target object of interest to characterize its environment (Section\ \ref{overdensity.sec}). Next we clean up our samples by removing foreground stars based on an observed-frame colour-colour cut in the $J-K_s$ vs. $u-J$ colour space (\citealt{Muzzin2013}). The objects are classified as galaxies if they satisfy the following criteria:
\begin{equation}
\begin{array}{l}
     \rm{SG2DPHOT} = 0, \\
     J-K_s>0.18 \times (u-J)-0.1, ~{\rm{if}}~ u-J<4.0,  \\
     J-K_s> -0.1, ~{\rm{if}}~ u-J\geq4.0, \\
\end{array}
\end{equation}
where SG2DPHOT is the KiDS-CAT star/galaxy classification bitmap based on the $r$-band source morphology (see, e.g., \citealt{deJong2015}). Values of the SG2DPHOT keyword have the following meaning: 1 = high confidence star candidate; 2 = unreliable source (e.g. cosmic ray); 4 = star according to star/galaxy separation criteria; 0 = all other sources (e.g. including galaxies). In order to ensure high-quality photometric redshifts, we require full 9-band photometry ($ugrizYJHK_s$) to be available for the galaxies in our sample. The total number of galaxies thus obtained in the KiDS+VIKING+HSC-SSP sample is 2,731,511 over the $\log(M_*[M_\odot])=9 - 11.5$ and $0<z<0.9$ range, and the number of quiescent galaxies (defined in Section\ \ref{type.sec}) among them is 478,677.  The choice to restrict our analysis to the $0<z<0.9$ range stems in part from our desire to ensure robust structural parameters of the stellar distribution based on rest-optical imaging.  Additionally, it is over the same redshift range that the $u$-to-$K_s$ photometry enables us to derive robust and well-calibrated photometric redshifts, while completeness effects, although present over some part of the redshift - mass parameter space (Section\ \ref{completeness.sec}), do not severely impact our analysis.

\subsection{Redshifts, stellar mass and galaxy type}
\label{z_Mstar_type.sec}

\subsubsection{Photometric redshifts}
\label{redshift.sec}

We use EAZY by \citet{Brammer2008} to determine the photometric redshifts, or use the spectroscopic redshift when available (albeit only for 3\% of all galaxies, and 9\% of the quiescent subset).\footnote{A compilation of spectroscopic redshifts is provided as part of the HSC-SSP public data release, containing amongst others many that were obtained by the GAMA spectroscopic survey \citep{Liske2015}.} The basic algorithm is to find the best-fitting non-negative linear combination of redshifted template spectra by using least-squares minimization.  The code includes an apparent magnitude prior on redshift (disfavouring the probability of finding extremely bright galaxies at very high redshifts, but equally accounting for the smaller volume probed at very low redshifts) and a wavelength-dependent template error function (such that more template mismatch is tolerated at the longest and shortest rest-wavelengths).  Minor zero point offsets for the observed photometric bands were determined by running EAZY on the spectroscopic sample with the redshift fixed to its spectroscopically determined value and evaluating the median residuals, a technique commonly adopted to fine-tune the photometric redshift quality (see, e.g., \citealt{Whitaker2011}; \citealt{Muzzin2013}).  The normalized median absolute deviation $\sigma_{\rm NMAD}$ of $\Delta z=z_{\rm{phot}}-z_{\rm{spec}}$ is calculated for the spectroscopic sample to quantify the quality of EAZY photometric redshifts:
\begin{equation}
    \sigma_{\rm{NMAD}} = 1.48\times \rm{median}(|\frac{\Delta z}{1+z_{\rm {spec}}}|),
\end{equation}
where the normalization factor of 1.48 ensures the standard deviation is retrieved in the case of a Gaussian distribution.
For the full sample of 76,607 spectroscopically confirmed galaxies with full 9-band photometry available, $0 < z_{\rm spec} < 0.9$ and $9 < \log(M_*) < 11.5$, the EAZY photometric redshifts are encouragingly consistent with the spectroscopic redshifts ($\sigma_{\rm{NMAD}}=0.026$).  For the subset of 41,336 quiescent galaxies, a similar quality ($\sigma_{\rm{NMAD}} = 0.025$) was found.  Where rest-frame colours are adopted, these are derived by the same code, by interpolating between the observed bands straddling the rest-frame wavelength of interest using the EAZY templates as a guide.

\subsubsection{Stellar mass}
\label{Mstar.sec}

In order to derive stellar population properties such as the galaxy stellar mass and star formation rate, we subject the 9-band spectral energy distributions (SEDs) to standard stellar population modelling techniques following the procedures outlined by \cite{Wuyts2011}.

Briefly, we fit \cite{Bruzual2003} models to the 9 band $u$-to-$K_s$ SEDs and search for the least-squares solution using the fitting code FAST (\citealt{Kriek2009}).
Here we apply the following constraints:
we allow ages since the onset of star formation between 50 Myr
and the age of the Universe, and visual (rest-frame $V$-band) extinctions in the range $0 < A_V < 4$ with the reddening following a \citet{Calzetti2000} law. We adopt a \citet{Chabrier2003} initial mass function (IMF), assume a uniform Solar metallicity, and allow exponentially declining SFHs with e-folding times down to $\tau=$ 300 Myr.

Stellar mass is one of the most robust parameters that can be extracted via stellar population synthesis modelling \citep[e.g.,][]{conroy2013, Mobasher2015}, and both its random uncertainties and systematics (e.g., due to the adopted family of star formation histories) remain well below the bin size for stellar mass we will adopt in this paper.

\subsubsection{Star-forming vs quiescent types}
\label{type.sec}

Star-forming galaxies (SFGs) are separated from quiescent galaxies (QGs) using a redshift-dependent cut in specific star formation rate: $\frac{\rm SFR}{M_*} > \frac{1}{3\  t_{\rm H}(z)}$, where $t_{\rm H}(z)$ represents the Hubble time at the observed redshift of the galaxy under consideration. We verified that the resulting classification is highly consistent (at the $\sim 90 - 98\%$ agreement level) with the one from selection wedges in rest-frame colour-colour space, such as the $V-J$ vs. $U-V$ (\citealt{Whitaker2012}) or $r-z$ vs. $u-r$ (\citealt{Holden2012}) diagrams.  We confirmed that a change between these different classification methods does not alter any of the conclusions presented in this paper.

The distribution of attenuation values inferred from our stellar population modelling are markedly different for QGs compared to SFGs, with the former peaking at very low $A_V$, whereas the $A_V$ distribution for SFGs extends to higher values and also in its mode shifts from $A_V \sim 0$ to $A_V \sim 1$ once we consider more massive galaxies ($\log M_* > 10$).  This is not surprising, as the richer interstellar medium content of SFGs compared to QGs is well established, but it is worth noting as a reassuring factor that attenuation effects are unlikely to bias our analysis of intrinsic 3D shapes for the QG population presented in this paper.  Conversely, we note in passing that the inclination-dependent attenuation seen in SFGs can usefully be exploited to place constraints on attenuation law shapes and dust geometries \citep[see, e.g.,][]{Wild2011}.

\begin{figure}
\centering
\includegraphics[width=\linewidth]{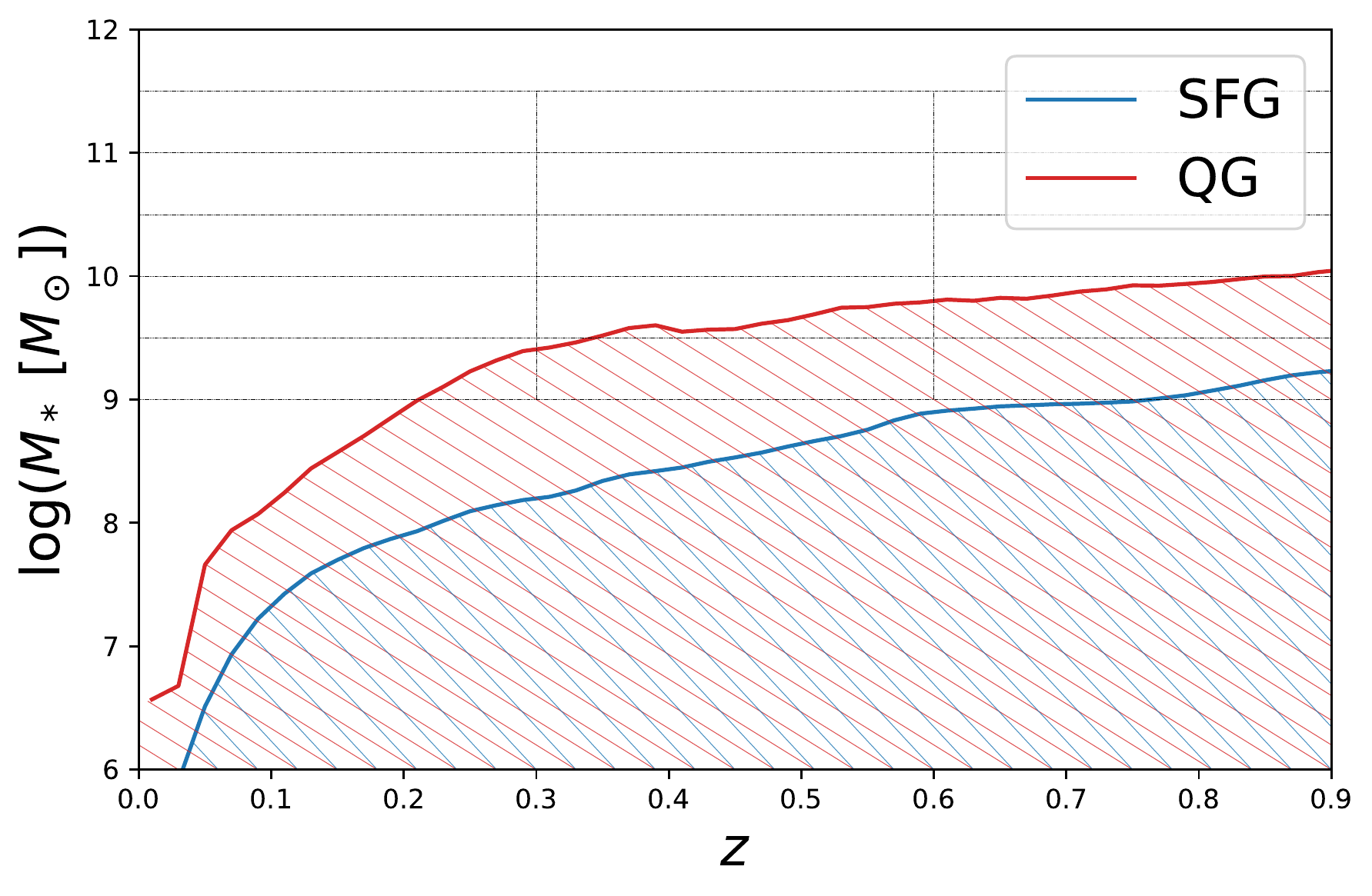}
\caption{
Stellar mass completeness limit for star-forming and quiescent galaxies as a function of redshift.  The grid of thin black lines denote the mass and redshift bins considered in this paper.
}
\label{fig:completeness}
\end{figure}

\subsubsection{Completeness}
\label{completeness.sec}

In order to assess the stellar mass completeness of our sample extracted from the KiDS+VIKING $ugrizYJHK_s$ 9-band joint catalogue, we consider for each galaxy by what factor the normalization of its SED could be reduced until it no longer meets the criteria to enter the sample.  Its stellar mass reduced by the same factor then represents the limiting mass corresponding to the faintest galaxy of this SED type that could have made it into our sample.  For a given galaxy type (SFG/QG) and redshift, we then determine the 90\% completeness limit as the 10th percentile of these limiting mass values.

Specifically, the criteria to enter our sample comprise a $S/N > 5$ in the SExtractor \citep{Bertin1996} FLUX\_AUTO parameter as quantified from the $r$-band detection image, and a $S/N > 1$ in all of the 9 bands as otherwise no numerical value for the photometry in the respective band would be included in the KiDS+VIKING catalogue.  A $z_{\rm phot} - z_{\rm spec}$ comparison for sources with only partial photometry included in the KiDS+VIKING catalogue confirmed their reduced photometric redshift quality, prompting us to define our analysis sample by the requirement to have 9-band photometry (albeit not necessarily a statistically significant detection in all of the bands).

Following the above procedure, we empirically derive the stellar mass completeness curves presented in Figure\ \ref{fig:completeness}, computed for SFGs (in blue) and QGs (in red) separately.  SFGs have a relatively lower detection limit at a given redshift compared to QGs, as anticipated from their younger stellar population and thus lower stellar mass-to-light ($M/L$) ratio.  We note that dust attenuation, where present, affects $M/L$ ratios as well, and our empirical completeness determination implicitly includes this effect.  The grid of three redshift intervals and stellar mass bins of 0.5 dex width marked in Figure\ \ref{fig:completeness} corresponds to the subsamples of galaxies as we will consider them in various parts of our analysis.  Those bins that may potentially be affected by incompleteness effects will always be explicitly marked as such.  We note though that, when it comes to the characterization of intrinsic shapes, incompleteness comes only into play if those QGs that failed passing the detection threshold and selection criteria are either of a different intrinsic shape than those in the same redshift and mass bin that entered the sample, or if they are preferentially seen under certain viewing angles.  The latter is an effect unlikely to apply to QGs which tend to be dust-poor and optically thin, thus not suffering from inclination-dependent attenuation effects.

\subsection{TNG simulations}
\label{TNG.sec}

Simulations offer a complementary perspective on galaxy evolution.  Unlike observations, they retain full knowledge of the 3D spatial distribution and each galaxy's star formation and assembly history. In this work we make use of IllustrisTNG, which is a state-of-the-art suite of large volume, cosmological, gravo-magnetohydrodynamical simulations including a comprehensive model for galaxy formation physics \citep{Marinacci2018, Naiman2018, Nelson2018, Pillepich2018a, Springel2018}. Each TNG simulation self-consistently solves for the coupled evolution of dark matter, cosmic gas, luminous stars, and supermassive blackholes from early times to the present day ($z=0$), allowing us to follow the formation and evolution of galaxies in its volume across cosmic time \citep{Weinberger2017, Pillepich2018b}. TNG resolves all but the lowest mass galaxies while simultaneously preserving the context of the larger cosmological volume.

The TNG project is made up of three simulation volumes: TNG50, TNG100, and TNG300 with cubic volumes of roughly 50, 100, and 300 Mpc side length. In our work, we make use by default of the public release of the TNG100 simulation, as it provides a balance between volume (and thus number statistics, including on the massive end) on the one hand, and spatial resolution on the other hand (see \citealt{Pillepich2019} \S2.3 for a discussion of numerical resolution in terms of particle masses, gravitational softening lengths and the range of gas cell sizes over which the magnetohydrodynamics is solved).  We further made use of the smaller volume, yet higher spatial resolution TNG50 simulation (\citealt{Nelson2019b}; typical cell sizes of 70 - 140 pc compared to 190 - 355 pc for TNG100) as a means to verify convergence of the intrinsic shape measurements and population trends based thereupon.  We study the intrinsic shapes of TNG simulated galaxies alongside the observations from KiDS+VIKING+HSC-SSP  by dividing them according to the same mass bins, applying the same QG selection criterion as outlined in Section\ \ref{type.sec}.  Within the context of the simulation, full knowledge of the galaxies' assembly history is available and intrinsic shapes can be determined on an individual object basis, not just statistically.

For full details on the IllustrisTNG public data release, we refer the reader to \citet{Nelson2019a}.  For a TNG50 based study of intrinsic shapes, specifically focussed on SFGs, we refer the reader to \citet{Pillepich2019}.

\section{Methodology}
\label{methodology.sec}

\subsection{3D intrinsic shapes}
\label{shapes.sec}

\subsubsection{3D shapes from observations}
\label{shape_obs.sec}

\begin{figure}
\centering
\includegraphics[width=\linewidth]{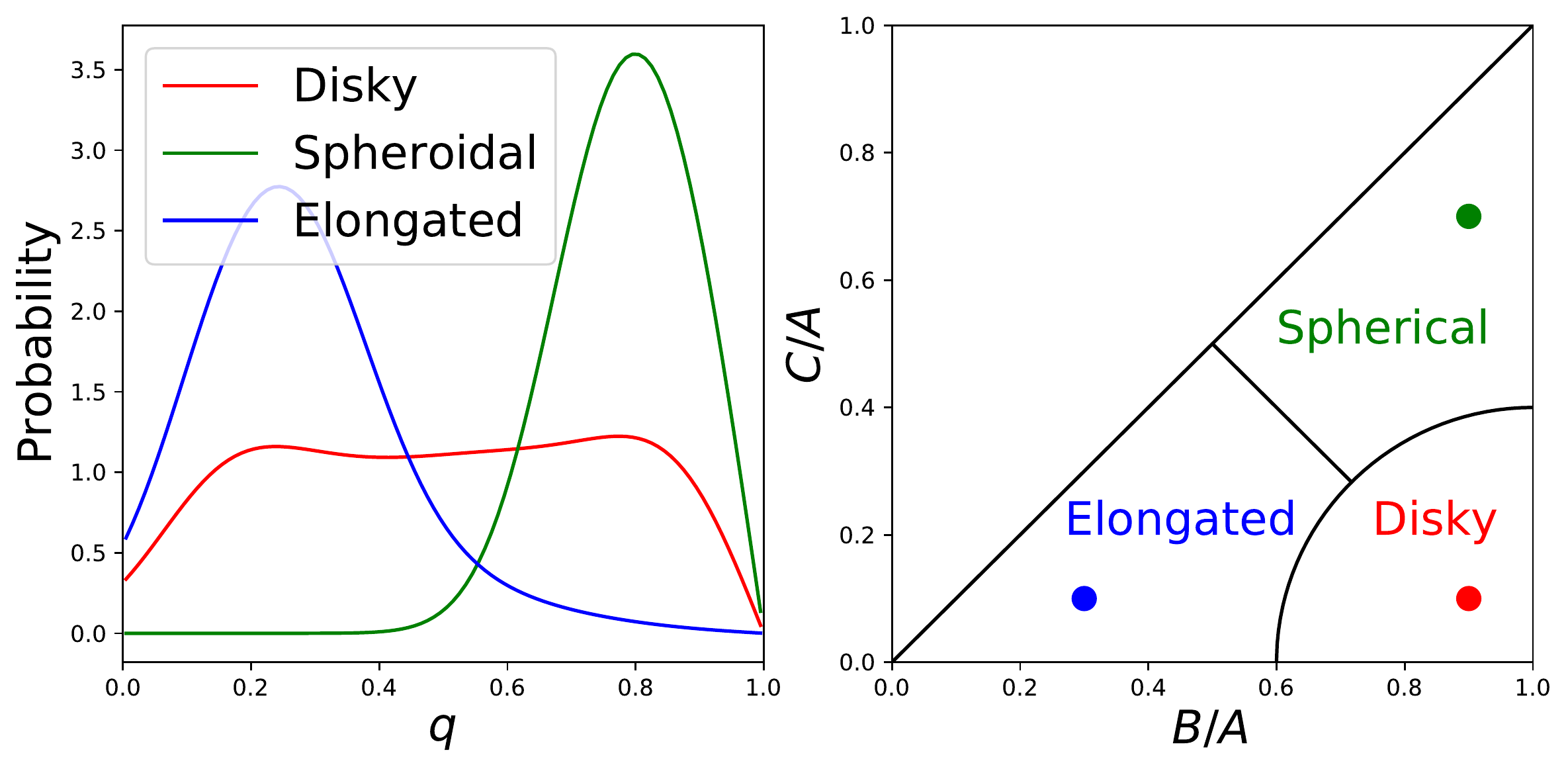}
\caption{
{\it Left panel:} Examples of the projected axial ratio distribution of a disky (red), spheroidal (green) and elongated (blue) galaxy as seen from random viewing angles.  {\it Right panel:} Corresponding position of the three case examples in intrinsic shape space, where C/A denotes the ratio of intrinsic minor to major axis length and B/A the ratio of intermediate to major axis length.  Where fractions of disky, spheroidal or elongated shapes are quoted, these are defined by the regions outlined in black.
}
\label{fig:shape}
\end{figure}

We follow the statistical method outlined by \citet{van2014b} to derive the 3D intrinsic shape of a galaxy ensemble from their distribution of 2D projected shapes. The galaxy ensemble is created by splitting our observed galaxy sample according to their internal properties such as redshift and stellar mass. The galaxy shape is simplified as a triaxial ellipsoid with intrinsic semi-axis lengths $A>B>C$. Let us call $\beta=B/A$ and $\gamma=C/A$. We follow
\cite{van2014b} in defining three types of galaxy shapes (disky, elongated, spheroidal) based on their position in the triangle plot of $C/A$ versus $B/A$ (Figure \ref{fig:shape}). The extreme case of an infinitely thin axisymmetric disk has $\beta=1$ and $\gamma=0$. Elongated galaxies are characterised by intrinsic axial ratios closer to $\beta=0$ and $\gamma=0$. Finally, a spherically symmetric galaxy would be described by $\beta=\gamma=1$. Figure \ref{fig:shape} shows an example of the intrinsic shape of disky, elongated, spheroidal galaxies and their projected axis ratio distribution when observed from uniformly distributed random viewing angles. In the projected axis ratio distribution, these three types of galaxies show significantly distinct features. Disky galaxies have a smooth, broadly extended distribution. Spheroidal galaxies feature a peak at high values of the projected axial ratio $q=b/a$. In contrast, elongated galaxies have a low-value peak.

If we project the triaxial ellipsoid with polar $\theta$ and azimuthal $\phi$ viewing angles in a
spherical coordinate system, we get the projected axis ratio $q$ as follows (see \citealt{Chang2013} Equation (3a) to (3d)).
\begin{equation}
    P_1 = \frac{\cos^2 \theta}{\gamma^2}(\sin^2 \phi+\frac{\cos^2 \phi}{\beta^2})+\frac{\sin^2 \theta}{\beta^2}
\end{equation}
\begin{equation}
    P_2 = \cos \theta \sin 2\phi (1-\frac{1}{\beta^2})\frac{1}{\gamma^2}
\end{equation}
\begin{equation}
    P_3 = (\frac{\sin^2 \phi}{\beta^2}+\cos^2 \phi)\frac{1}{\gamma^2}
\end{equation}
\begin{equation}
    q(\theta,\phi;\beta,\gamma)=\sqrt{\frac{P_1+P3-\sqrt{(P_1-P_3)^2+P_2^2}}{P_1+P3+\sqrt{(P_1-P_3)^2+P_2^2}}}
\end{equation}

In reality, we can only determine intrinsic 3D shapes from galaxies with similar structures but viewed at random orientations. To introduce a variance for galaxies with similar intrinsic shapes, we assume a Gaussian distribution of the ellipticity $E$ and the triaxiality $T$ with dispersion $\sigma_E$ and $\sigma_T$, where $E$ and $T$ are defined as:
\begin{equation}
    E=1-C/A
\end{equation}
\begin{equation}
    T=(A^2-B^2)/(A^2-C^2)
\end{equation}

On the other hand, from the projected images, we can derive galaxies' semi-major $a$ and semi-minor $b$ axis lengths by diagnosing the mass eigentensor Q.

\begin{equation}
    Q = \begin{bmatrix} 
          \cos(\theta) & \sin(\theta)\\
          -\sin(\theta) & \cos(\theta)
        \end{bmatrix}
        \begin{bmatrix}
          a^2 & 0 \\
          0   & b^2    
        \end{bmatrix}
        \begin{bmatrix}
         \cos(\theta) & -\sin(\theta)\\
         \sin(\theta) & \cos(\theta)
        \end{bmatrix}
\end{equation}
The projected axis ratio $q$ is:
\begin{equation}
    q = b/a
\end{equation}

For nearly round ($q \sim 1$) galaxies, random noise will always cause the measured $q$ to be an underestimate as the position angle of the long axis becomes ill-determined. This affects the projected axis-ratio distribution as described by \cite{Rix1995} (Equation (C5)):

\begin{equation}
    P_e(\epsilon|\epsilon_e,\Delta \epsilon)=\frac{\epsilon}{\Delta \epsilon^2}I_0(\frac{\epsilon\epsilon_e}{\Delta\epsilon^2})\exp(-\frac{\epsilon^2+\epsilon_e^2}{2\Delta\epsilon^2})
\end{equation}
where $\epsilon = 1- q$ is the measured ellipticity, $\epsilon_e$ is the expected ellipticity, $\Delta \epsilon$ is the measurement error, $P_e$ is the expected ellipticity distribution and $I_0$ is the modified Bessel function of the first kind with order zero. For the HSC-SSP data set, we apply a fixed value $\Delta\epsilon=\Delta q=0.1$, motivated by the comparison to $q$ measurements on CANDELS $H$-band imaging where available (Section\ \ref{KiDS_VIKING_HSC.sec}). Then for a given set of model parameters $E$, $T$, $\sigma_E$ , and $\sigma_T$, we can generate a model $q$ distribution that incorporates the effect of measurement errors, and evaluate the log likelihood of obtaining the observed $q$ distribution given the model as:
\begin{equation}
    L = \sum_{i} \log (P(n_i|m_i)).
\end{equation}
Here, for each bin $i$,  $n_i$ is the number of observed galaxies in that bin and $m_i$ is the number predicted by the model with specific parameters. The conditional probability can be approximated by a Poisson distribution:
\begin{equation}
    P(n_i|m_i) = e^{-m_i} m_i^{n_i} / n_i !
\end{equation}
The log likelihood thus becomes (\citealt{Holden2012}):
\begin{equation}
    L = \sum_{i} (-m_i + n_i\log(m_i) - \log(n_i!)) = \sum_{i} n_i\log(m_i) + C_1,
\end{equation}
where $C_1=\sum_{i} (-m_i - \log(n_i!)) = -N - \sum_i(\log(n_i!))$ is a constant for a fixed bin size and observed sample. In maximizing the likelihood function, we can therefore ignore the constant term $C1$ and only calculate the rest part. Besides, we can not guarantee $m_i$ is always larger than 0, so we impose a floor of 1 to avoid negative infinite values.
 \begin{equation}
    L' = \sum_{i} n_i \log(m_i+1).
\end{equation}

 In practice, we estimate the probability distribution function by drawing 10000 galaxies with random viewing angles. Then the predicted number of galaxies in each bin can be derived by multiplication with the ratio $N_{\rm{obs}}/10000$.  Finally, rather than modelling the one-dimensional $q$ distribution, we follow \citet{Zhang2019} in fitting the distribution of the ensemble of galaxies across the two-dimensional $q - \log(a)$ space, where $a$ represents the projected semi-major axis length.\footnote{We note that the projected semi-major axis length only equals the intrinsic (3D) semi-major axis length in the case of oblate axisymmetric systems.} 
 
 \cite{Zhang2019} based their approach on the finding that galaxies with smaller projected semi-major axis $a$ are rounder at all stellar masses and redshifts. Hence they included the projected semi-major axis $a$ as a second dimension in their modelling, with a covariance ${\rm{Cov}}(E,\log(a))$ to describe the relation between ellipticity and semi-major axis length. Their model can reduce systematic residual patterns compared to the fiducial $q$-only modelling. Throughout this paper, we therefore apply the $q$-$\log(a)$ model to deproject the intrinsic 3D shapes. In this case, we bin up the $q$-$\log(a)$ plane and the log likelihood becomes:
\begin{equation}
    L = \sum_{i,j} n_{i,j} \log(m_{i,j}).
\end{equation}
For each bin with index ${i,j}$,  $n_{i,j}$ is the number of galaxies in that bin and $m_{i,j}$ is the number predicted by a model defined by a specific set of parameter values.

 The number of bins adopted to divide this two-dimensional space ($N_{\rm{q, bin}}$, $N_{\rm{\log(a), bin}}$) is chosen to scale with the number of observed galaxies, in order to maintain a balance between sufficient resolution elements along each of the $q$ and $\log(a)$ dimensions while simultaneously ensuring robust number statistics within each bin:
 \begin{equation}
    \begin{array}{l}
          N_{\rm{q, bin}} = {\rm{max}}({\rm{min}}({\rm{int}}(2\sqrt{N_{\rm{obs}}/20}),50), 24)\\
          N_{\rm{\log(a), bin}} = {\rm{max}}({\rm{int}}(N_{\rm{q, bin}}/2),12)
    \end{array}
\end{equation}
where $N_{\rm obs}$ represents the total number of galaxies in the ensemble that is being modelled.  We further require the number of observed galaxies to be at least 100 to ensure robust results. In this case, on average we will have at least 1 galaxy in each $q-\log(a)$ bin and more bins (of smaller bin width, yielding a smoother distribution) for sub-samples with a larger number of objects. 
 
Given the projected $q - \log(a)$ distribution of an observed galaxy ensemble, we search for the best-fitting ellipticity $E$, triaxiality $T$ and their dispersion $\sigma_E$ and $\sigma_T$ by using the {\tt emcee} \citep{EMCEE} implementation of the Markov Chain Monte Carlo (MCMC) method.  In fact, upon experimentation we found there to be a statistically significant improvement in the ability to reproduce the observed $q - \log(a)$ distributions when allowing for two shape families ($E_1, \sigma_{E1}, T_1, \sigma_{T1}$) and ($E_2, \sigma_{E2}, T_2, \sigma_{T2}$), with an additional parameter $f$ quantifying the relative fraction belonging to each family.  When quoting single values of ellipticity and triaxiality, these will hereafter refer to the 50th percentile of the overall $E$ and $T$ distributions, respectively.

\subsubsection{3D shapes from simulations}
\label{shape_sim.sec}

Different methods can be applied to distill the 3D coordinates of individual stellar particles belonging to a simulated galaxy into single values of $C/A$ and $B/A$ approximating the galaxy's stellar distribution (see, e.g., \citealt{Pillepich2019}, who further document that significantly flatter structures are found when quantifying the intrinsic shapes of the star-forming gas rather than the stellar distribution within TNG galaxies).  In this section, we contrast three such methods to quantify the TNG galaxy intrinsic shapes.

The first method is measuring the mass eigentensor within two stellar half-mass radii ($2R_{\rm{half}}$).  These quantities are provided as part of the stellar circularity table included as one of the TNG project's default data release products (\citealt{Nelson2019a}; see also \citealt{Genel2015}).

The second method is similar to the first, but measures the mass eigentensor using only stellar particles in an ellipsoidal shell at $R_{\rm{half}}$ (\citealt{Pillepich2019}). It starts with the mass eigentensor matrix derived from a spherical shell at $R_{\rm{half}}$ (specifically considering all stellar particles within $R_{\rm half}\pm 0.2 R_{\rm half}$), and calculates the corresponding shape parameters with this matrix. With these shape parameters, a new ellipsoidal shell is subsequently defined to pick up the stellar particles within this shell. After several iterations, the shape parameters of the ellipsoidal shell converge. 

The third method calculates the minimum volume enclosing ellipsoid (MVEE) that covers all the particles in the system \citep{Nima2006}. The shape of the MVEE is strongly determined by the outermost particles. In order to obtain a shape measurement characteristic for the overall stellar distribution, we throw away the outermost particles in each iteration, and calculate the next MVEE with the remaining particles.  Doing so iteratively, we obtain a quantification of the intrinsic axial ratios as a function of enclosed mass.  To distill a single value of $C/A$ and $B/A$ for a given galaxy, we take the median over the range in enclosed stellar mass from 50\% to 80\%, where shape parameters are found to be well converged.\footnote{Taking the median over the 45\%-55\% range in enclosed stellar mass instead makes little difference, the mean absolute deviation between $C/A$ values computed based on the 50\%-80\% vs 45\%-55\% enclosed mass intervals is 0.034, and similarly small for $B/A$.} The MVEE method is time-consuming when the number of particles is large. \cite{VanAelst2009} proposed a means to simplify the system without loss of shape information, by resampling and combining the particles. If the system has $N$ particles marked as $x_i$, we want to simplify the system to $P$ particles. We can randomly pick up $N_1$ groups, each with $N_2$ particles with replacement and each group is marked as $X_i$:
\begin{equation}
    X_i = \{x_{i,1}, x_{i,2}, ..., x_{i,N_2}\},
\end{equation}
where $x_{i,j}$ are the particles randomly picked from the full sample. Then a set of new particles $y_i$ can be generated by averaging over the particles in group $X_i$:
\begin{equation}
    y_i = 1/N_2\times\sum_{j=1}^{N_2} x_{i,j}.
\end{equation}
In this way, one can generate a simplified system with $N_1$ particles to accelerate the speed to derive the shape, while simultaneously preserving as much information as possible to guarantee an accurate representation of the true shape.

\begin{figure}
\centering
\includegraphics[width=\linewidth]{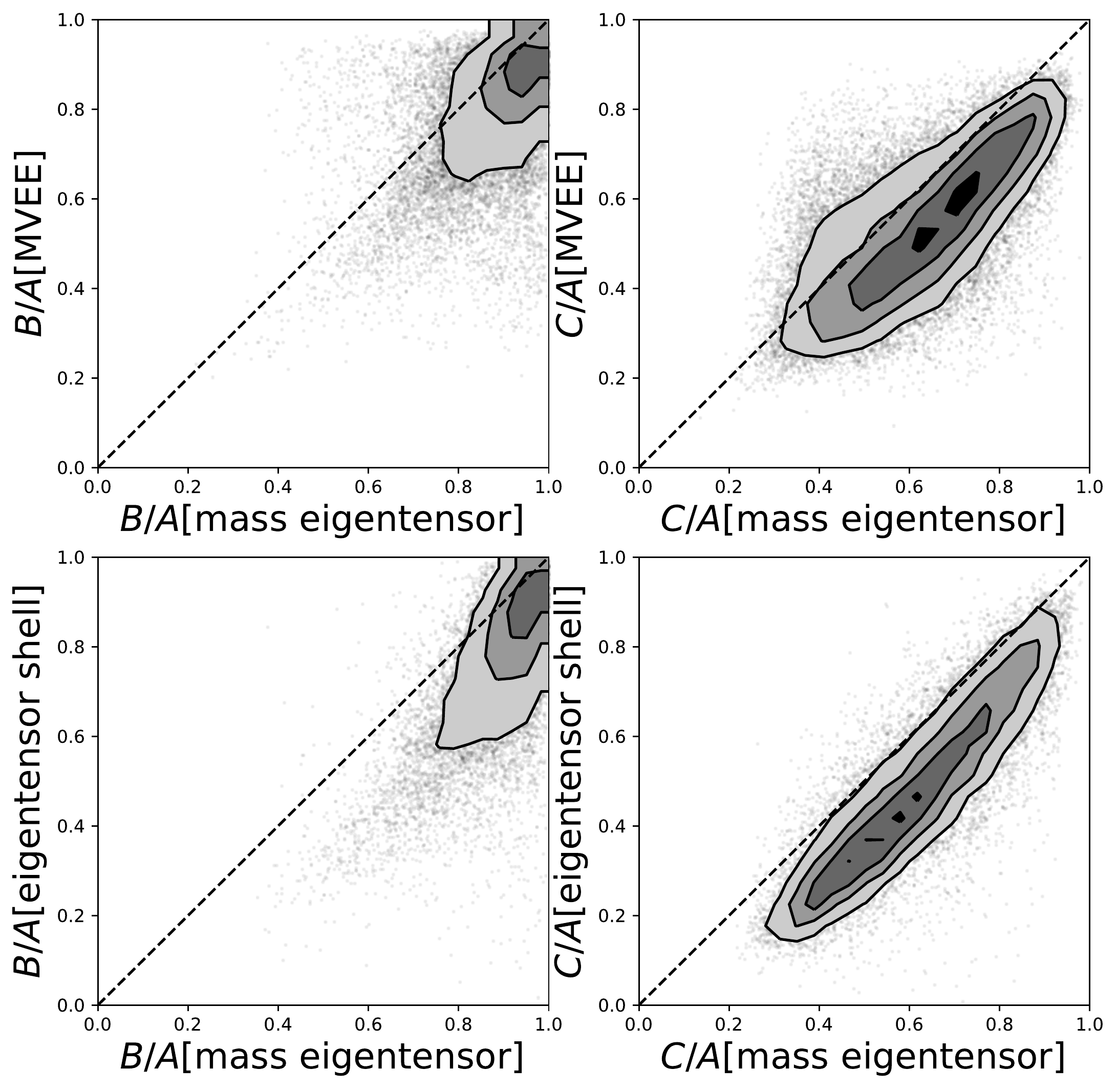}
\caption{
Comparison between three different methods to quantify the stellar intrinsic shape of simulated galaxies: the minimum volume enclosing ellipsoid (MVEE), diagonalization of the mass eigentensor adopting all stellar particles within $2R_{\rm star,half}$ or only those in a shell around $R_{\rm star,half}$.  Left- and right-hand panels contrast the intermediate-to-major and minor-to-major axis ratio, respectively.  See text for details on methods and the simulated galaxy population shown.  More flattened shapes are retrieved by methods that emphasize outer (or near-$R_{\rm star,half}$) shapes and ignore inner stellar particles.}
\label{fig:simshape}
\end{figure}

Figure \ref{fig:simshape} shows a quantitative comparison between the three methods. The galaxies included are extracted from TNG100, for the snapshot corresponding to redshift $z = 0.44$ and spanning a similar mass range as our observational analysis sample ($\log M_* > 9$).  A similar behaviour albeit with smaller number statistics is seen in TNG50.  Both the mass eigentensor shell and MVEE methods yield shape measurements that are less round than those obtained using the mass eigentensor method. Since the stellar shapes of TNG galaxies are in general rounder than those observed in the real universe, we refrain from using the default mass eigentensor method.  Relatively speaking, the other two methods place more emphasis on the outer distribution of particles, which is more akin to the situation in observed analyses where most constraining power on the axial ratio comes from larger radii, not the beamsmeared centre of the galaxies.
The mass eigentensor shell method is quite efficient but it is not reliable if the number of stars within the shell is less than 100. The MVEE method on the other hand is accurate if we do not over-simplify the system, but it requires too much computation when the galaxy contains too many stellar particles. So we combine the intrinsic shapes derived from the previous two methods by using the MVEE method for low-mass galaxies and the mass eigentensor shell method for high-mass galaxies.

\subsection{Environment}
\label{environment_method.sec}

The environment in which a galaxy lives can potentially be an important factor influencing a galaxy's shape.  This is certainly expected when considering the galaxy population as a whole (i.e., SFGs + QGs), as expressed in the morphology - density relation \citep{Dressler1980}. The fraction of early-type galaxies increases and late-type galaxies decreases
toward increasing local galaxy density (\citealt{Goto2003}). Various physical mechanisms that can contribute to such a relation are listed in Section\ \ref{sec:Intro}. 

In this paper, we focus specifically on the quiescent galaxy population, and will investigate whether among them an environmental impact on galaxy shapes is seen.  This makes for a question that is distinct from whether an overall morphology - density relation is present.  The latter is dominated by an environmentally dependent quenched fraction, whereas we intend to contrast the shapes of galaxies that are all quenched, but living in different environments (see also \citealt{PaulinoAfonso2019}).  Our methodology to characterize a galaxy's local environment is detailed in this section.

\subsubsection{Overdensity}
\label{overdensity.sec}

To quantify the environment, we firstly use the local galaxy number density $\rho$ which is calculated by counting the number of neighbouring galaxies $N_{\rm{Galaxy}}$ (of any type, SFG or QG) within a cylinder with a proper volume $V_{\rm{proper}}$ as follows
\begin{equation}
    \rho = N_{\rm{Galaxy}}/V_{\rm{proper}}.
\end{equation}
The proper volume is determined by a cylinder with a $0.25$ pMpc (proper Mpc) radius sky aperture and a depth determined by the distance along the line of sight between a redshift $z-\delta z$ and $z+\delta z$, where $z$ represents the target galaxy's redshift and.
\begin{equation}
    \delta z = (1+z)\times \sigma_{\rm NMAD}, 
\end{equation}
where $\sigma_{\rm NMAD}$ is the normalised median absolute deviation for KiDS+VIKING photometric redshifts as measured against a spectroscopically confirmed reference sample, as described above.
The proper volume can be defined as:
\begin{equation}
    V_{\rm{proper}}=V_{\rm{cosmos}}/(1+z)^3
\end{equation}
\begin{equation}
    V_{\rm{cosmos}} = \theta(z)^2/4\times(V_{\rm{comoving}}(z+dz)-V_{\rm{comoving}}(z-dz))
\end{equation}
where $V_{\rm{comoving}}(z)$ is the spherical volume enclosed within redshift $z$, and $\theta(z)$ refers to the opening angle spanned by a 0.25 Mpc physical (proper) transverse distance at the redshift z of the galaxy under consideration.
\begin{equation}
    \theta(z) = \rm{arcmin/pkpc}(z)\times 1000/60\times\pi/180
\end{equation}

In our case we do not have spectroscopic redshifts for all galaxies in our sample, so we can not guarantee that the neighbouring galaxies are located inside the cylinder. Based on the photometric redshifts and the $\pm 1\sigma$ error on $z_{\rm phot}$ in hand for each object, we reconstruct a redshift probability distribution $P_i(z)$ composed of two half-Gaussians for a certain galaxy $i$. We calculate the probability that the galaxy is located within the cylinder centered on our target galaxy at redshift $z$ as:
\begin{equation}
    P_i = \int_{z-\delta z}^{z+\delta z} P_i(z) dz
\end{equation}
Then the weighted number of galaxies within the cylinder is:
\begin{equation}
    N_{\rm{Galaxy}} = \sum_{i}P_i
\end{equation}
Where the circular aperture on the sky around a target galaxy overlaps with a sky region masked due to the vicinity of a bright foreground star, this is account for by adjusting $V_{\rm proper}$ downward accordingly.  If more than half of the sky aperture is masked, we consider the overdensity measure unreliable and simply eliminate the galaxy from any analysis that involves environment.  We emphasize that this does not introduce any bias to our analysis as the masked regions are due to bright foreground stars, unrelated to the galaxies we study.

In TNG, the 3D positions of all galaxies are known. In this case, we determine the number density within a $0.25$ pMpc radius sphere at the target galaxy's location:
\begin{equation}
    \rho = N_{\rm{Galaxy}}(<{\rm{0.25pMpc}})/V_{\rm{proper}}(<{\rm{0.25pMpc}})
\end{equation}

To obtain a measure of relative overdensity indicating how the local density is larger or smaller than the mean value, we compute the average density for galaxies within the same redshift bin and use the local overdensity $\delta\rho$:
\begin{equation}
    \delta\rho=(\rho-\overline{\rho})/\overline{\rho}
\end{equation}

\begin{figure}
\centering
\includegraphics[width=\linewidth]{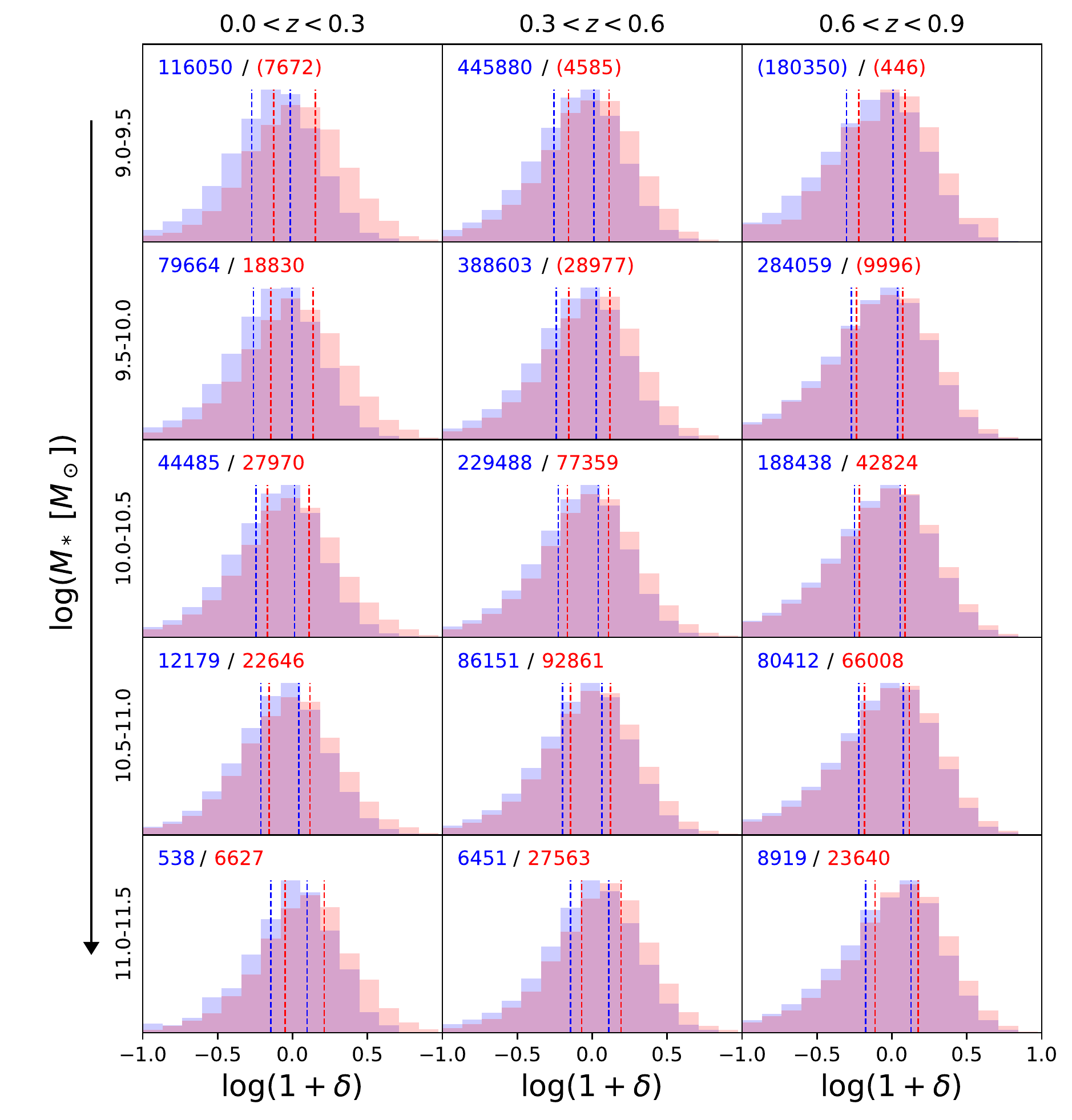}
\caption{
Overdensity distributions of QGs (red) in bins of stellar mass and redshift, with SFGs shown in blue for reference.  Vertical dotted lines mark tertiles of the distributions and inset numbers in the upper left of each panel denote the sample size (in brackets where partially incomplete).
}
\label{fig:overdensity}
\end{figure}

The distribution of environmental overdensities quantified for the galaxies in our observed sample is shown in Figure \ref{fig:overdensity}, split by mass and redshift. Blue and red histograms illustrate separately the overdensity distribution of star-forming and quiescent galaxies with vertical dotted lines marking lower and upper tertiles.  It is evident that, while there is significant overlap between the distributions, the overdensity distribution of QGs is systematically offset with respect to that of SFGs.  QGs live, on average, in denser environment than SFGs and their relative fraction increases as one considers more extreme overdense regions.  This trend is most pronounced at lower redshifts and lower masses, and its very presence illustrates that environmental effects are not fully washed out at the photometric redshift quality reached on the basis of broad-band photometry alone.  That said, from the relative number of SFGs and QGs quoted in each panel of Figure\ \ref{fig:overdensity}, it can be appreciated that the quenched fraction is a much stronger function of galaxy stellar mass than it is of environment.

\subsubsection{Tidal parameter}
\label{tidal.sec}

When we further consider the interaction between galaxies and their neighbours, tidal stretching can be a mechanism to affect their shapes. Quantifying the potential impact by tidal forces requires more parameters than the local overdensity (e.g., the size of the target galaxy and masses of its neighbours). We apply a tidal parameter proportional to  the tidal force over the binding force $F_{\rm{tidal}} / F_{\rm{bind}}$ as follows:
\begin{equation}
    Q_{i,p} = (M_i/M_p)*(A_p/R_{i,p})^3
    \label{eq:Q_pi}
\end{equation}
where $Q_{i,p}$ is the tidal parameter for primary galaxy $p$ induced by the neighbouring galaxy $i$. $M_p$ and $A_p$ are respectively the stellar mass and projected size of galaxy $p$. Likewise $M_i$ and $R_{i,p}$ are respectively the stellar mass of galaxy $i$ and the projected distance between galaxy $i$ and galaxy $p$. The total tidal parameter for a given galaxy is the obtained by summing over the tidal impact felt from all of its neighbours:
\begin{equation}
    Q_{p} = \sum_{i}Q_{i,p}
    \label{eq:Q_p}
\end{equation}
Note that, when applied to observations, we do not have the distance information in 3D, so we use the projected ones. Likewise, the projected size of the target only equals its intrinsic major axis size in the case of oblate axisymmetric systems. However, when carrying out an equivalent analysis on simulated galaxies from the TNG simulation, we have the luxury of measuring these quantities in 3D as well as in projection, allowing us to evaluate the impact of projection effects.
In our case we do not have spectroscopic redshifts for the majority of galaxies, so we do not necessarily know for certain which are the neighbours that are physically associated. So we use a weight factor $P_{i,p}(z)$ between 0 and 1 that quantifies the probability that the neighbour $i$ is physically associated with the primary $p$ defined by the overlap between the two galaxies' redshift probability distribution functions:
\begin{equation}
    P_{i,p}(z) = \sum_{N_{\rm{bins}}} \sqrt{(P_i(z)*P_p(z))}
\end{equation}
where $N_{\rm{bins}}$ is the number of bins in the normalized probability distribution of the primary galaxy's redshift $P_p(z)$ and the neighbouring galaxies' redshift $P_i(z)$. 

Then we can adjust the Eq.\ \ref{eq:Q_p} to:
\begin{equation}
    Q_{p} = \sum_{i}Q_{i,p}*P_{i,p}(z)
\end{equation}

\begin{figure}
\centering
\includegraphics[width=\linewidth]{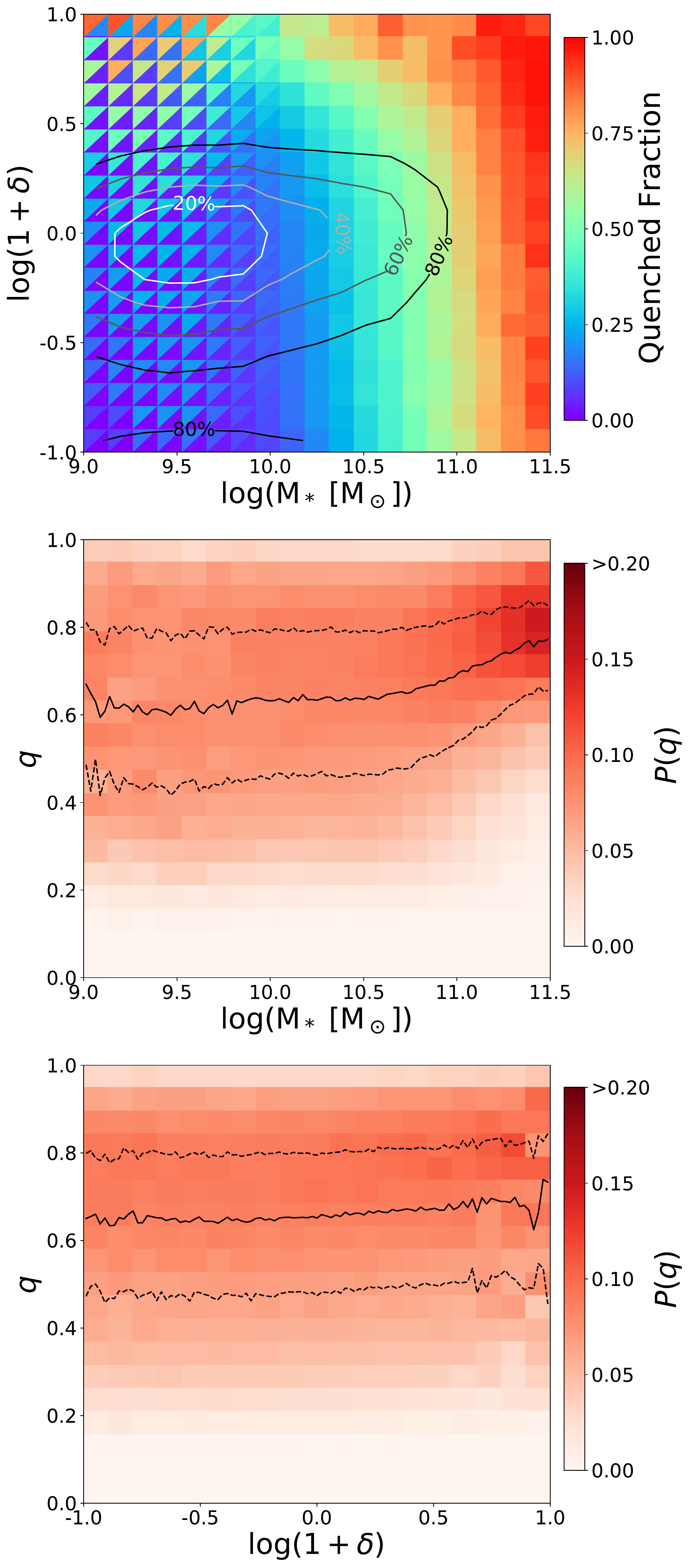}
\caption{
{\it Top:} Dependence of quenched fraction on stellar mass and overdensity.  White to black contours depict regions comprising an increasing percentage of galaxies in our $0<z<0.9$ sample.  Colour coding of the upper/lower triangles mark upper/lower limits on the quenched fraction, respectively (see text).  Larger quenched fractions are found at high mass or in overdense environments.
{\it Middle and bottom panels:} Running median and 25th/75th percentile curves for the projected axial ratio $q$ of QGs as a function of stellar mass and overdensity.  Shades of red depicts the probability density function of $q$ (normalized per column).  At $\log(M_*)>11$, QGs span a narrow range of high $q$ values (round in projection).  Any environmental dependence of the $q$ distribution is comparatively marginal.}
\label{fig:obs_fquench_q}
\end{figure}

For TNG, we similarly use Eq.\ \ref{eq:Q_pi} to calculate the tidal parameter. But since we know the 3D position of all galaxies, we are able to calculate the separation distance $R_{i,p}$ in 3D.

\section{Results}
\label{results.sec}

\subsection{Quenched fraction and axial ratios}

Before we delve into a detailed analysis of inferred intrinsic shapes controlling for mass and redshift (Section\ \ref{high_mass.sec}), and additionally environment (Section\ \ref{environment.sec}), we first summarize the overall quenched fraction and projected axial ratios as a function of galaxy mass and environmental density in Figure \ref{fig:obs_fquench_q}. Here, we combine the full $0<z<0.9$ redshift range, and in the case of projected axial ratios consider the dependence on one parameter at a time, i.e., marginalizing over all environments when considering the dependence on mass and vice versa.

Since quiescent galaxies have larger mass-to-light ratios and for the same mass are therefore more difficult to detect in a flux-limited survey, our sample of quiescent galaxies can be more incomplete than that of star-forming galaxies. The quenched fraction $N_{\rm{QG}}/N_{\rm{total}}$ can thus be underestimated if we use the exact number of observed quiescent galaxies and all galaxies. For this reason, we define this observed ratio as a lower limit on the quenched fraction. To make the results more reliable, we also estimate an upper limit to the quenched fraction. Considering the stellar mass completeness curve (Figure \ref{fig:completeness}), we can learn about the maximum redshift $z_{\rm{max}}(M_*, {\rm type})$ out to which a galaxy of certain mass and type (SFG/QG) would remain detectable and present in our sample. In other words, the number of such galaxies observed within the co-moving volume $V_c(z < z_{\rm max}(M_*,\ {\rm type}))$ would be complete, but outside $z_{\rm max}(M_*,\ {\rm type})$ galaxies would be missing from our sample.  Assuming (incorrectly) a quenched fraction that is independent of cosmic time over the $z$ range considered, a $1/V_{\rm max}$ completeness correction can then be applied where in quantifying the number of quiescent and all galaxies, each galaxy is assigned a weight:
\begin{equation}
    w = {\rm{max}}(V_c(z=0.9)/V_c(z_{\rm{max}}(M_*, {\rm{type}})),1).
\end{equation}
In reality, the quenched fraction is a monotonically increasing function of cosmic time, implying that the $1/V_{\rm max}$ correction applied to the broad $0<z<0.9$ redshift range will overcorrect for the missing lower mass QGs at higher redshifts, thus rendering the estimate to be an upper limit to the true quenched fraction.  Figure\ \ref{fig:obs_fquench_q} illustrates the above bracketing scenarios by adopting the lower/upper limits on quenched fraction as colour-coding for the lower/upper triangles of which each bin is composed.

The overall quenched fraction exhibits dependencies on both mass and overdensity.  Specifically, the top panel of Figure\ \ref{fig:obs_fquench_q} iterates results from local and smaller area intermediate redshift spectroscopic surveys by \citet{Peng2010}, reporting on the separability of mass and environment quenching.  Overall, quenched fractions depend most strongly on galaxy stellar mass.  However, at intermediate to low masses secondary effects are observed such that at fixed mass the quenched fraction increases with overdensity.

Turning to the middle and bottom panels of Figure\ \ref{fig:obs_fquench_q}, which display the $q$ distribution of QGs as a function of mass and overdensity, respectively, the magnitude of any systematic variation with these parameters reduces in that order.  The distribution of $q$ values remains approximately constant over 2 orders of magnitude in stellar mass, but then changes abruptly above $\log(M_*) > 11$, where they are observed to cluster around high values (i.e., round in projection).  This result echoes findings by \citet{van2009} based on nearby SDSS galaxies and by \citet{Holden2012} based on 1,332 QGs at $0.6 < z < 0.8$ extracted from the GEMS \citep{Rix2004} and COSMOS \citep{Scoville2007} surveys (see also \citealt{Satoh2019}, who exploit a sample of 8,162 passive or quenching galaxies out to $z=1$, still a factor $>50\times$ smaller than the sample considered here).  Comparatively, any systematic variations with overdensity are less pronounced.

In the following sections, we proceed to slice up parameter space more finely, controlling for both redshift and mass.  This enables treating the selected subsamples as ensembles that more plausibly can be considered to comprise objects of a common intrinsic shape viewed from random viewing angles.  Other than solving the inversion problem, we will also relate to results from the TNG simulation where appropriate.

\subsection{Spheroidal shapes at the high-mass end}
\label{high_mass.sec}

\begin{figure}
\centering
\includegraphics[width=\linewidth]{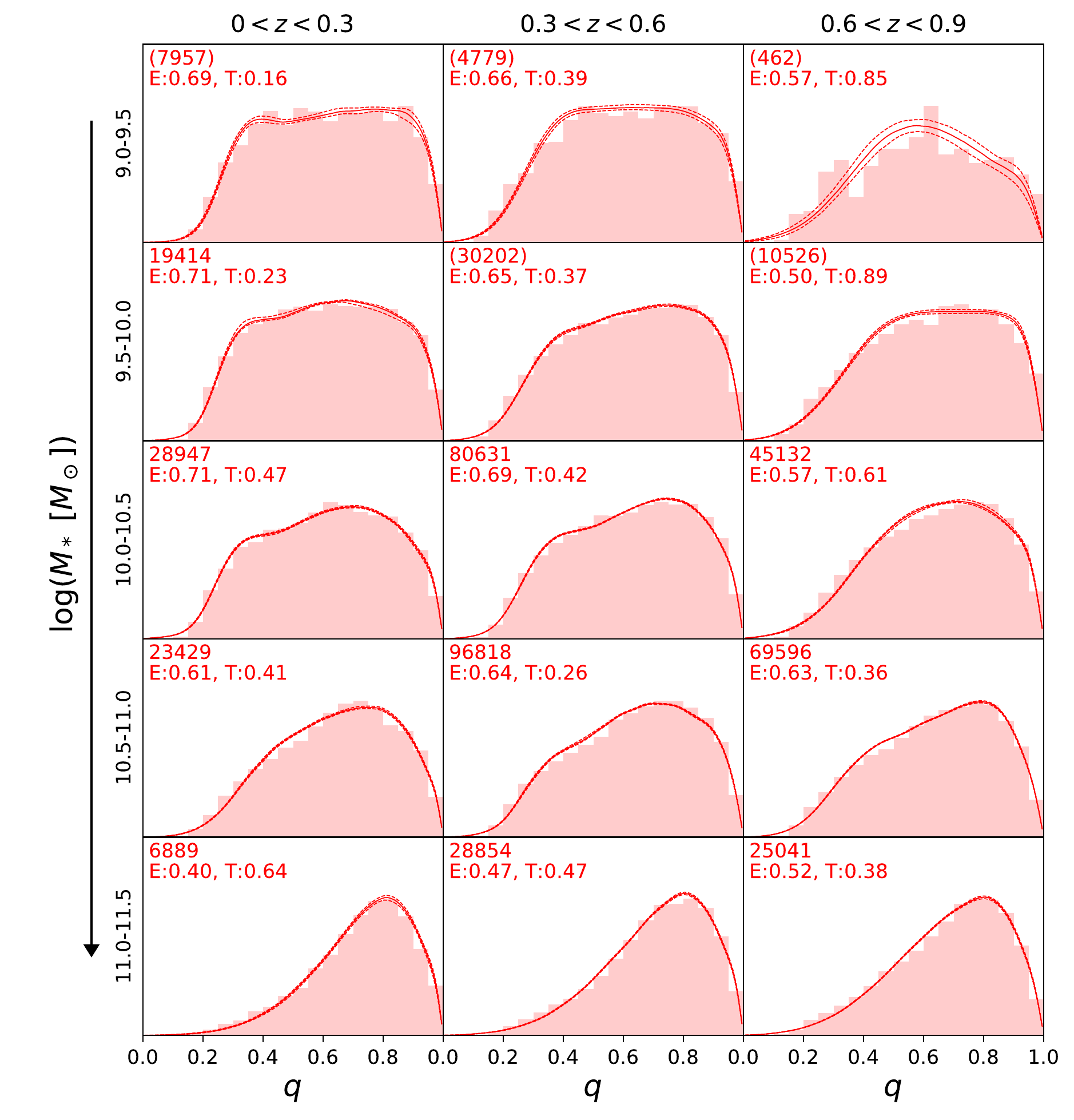}
\includegraphics[width=\linewidth]{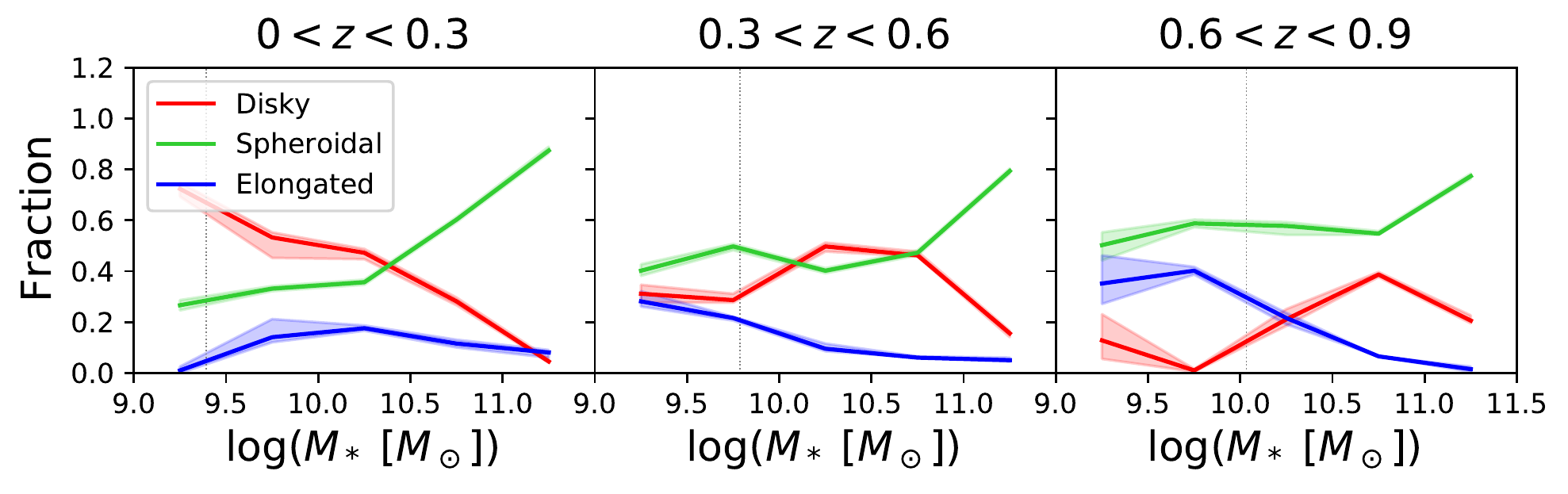}
\caption{{\it Top:} Observed distribution of projected axial ratios $q$ of QGs in bins of different mass and redshift. In the upper left of each panel, we mark the sample size (in brackets where partially incomplete) in the first row and the median value of the inferred ellipticity ($E$) and triaxiality ($T$) distributions in the second row. Associated 2D histograms of the $q - \log(a)$ distributions are depicted in Appendix\ \ref{appendixA.sec}.
{\it Bottom:} Inferred fraction of QGs with disky, elongated and spheroidal intrinsic shapes as a function of mass and redshift.  Dotted vertical lines mark the mass completeness limit for QGs at the upper end of the redshift interval.  At all redshifts a significant upturn in the fraction of intrinsically round systems is notable among the most massive QGs.
}
\label{fig:obs_shape_allQG}
\end{figure}

\begin{figure}
\centering

\includegraphics[width=\linewidth]{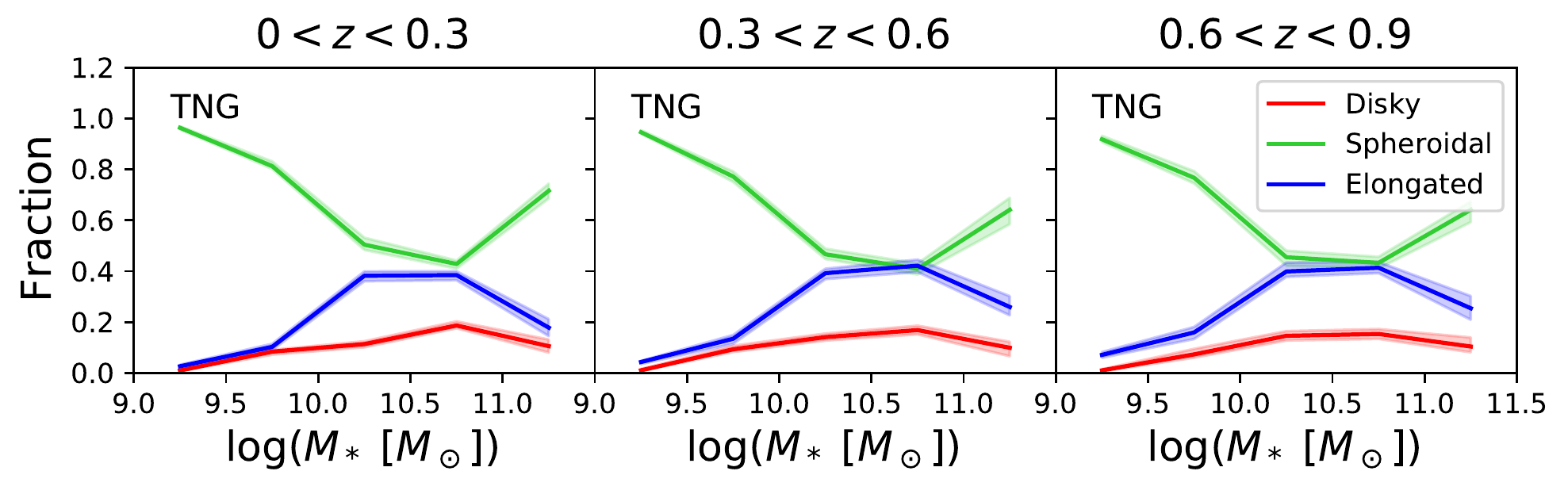}
\caption{Inferred fraction of simulated QGs in TNG with disky, elongated and spheroidal intrinsic shapes as a function of mass and redshift. Uncertainties are calculated by bootstrapping the sample with replacement.
}
\label{fig:sim_shape_allQG}
\end{figure}

We find spheroidal shapes to be dominant among quiescent galaxies with more than $10^{11}M_\odot$. The top grid of panels in Figure\ \ref{fig:obs_shape_allQG} splitting the quiescent galaxy population by mass and redshift, illustrate this in terms of the observed projected axial ratio distributions.  Here, filled histograms show the observed $q$ distribution whereas the best-fit model, optimized to reproduce the $q$ - $\log(a)$ distribution shown in Appendix\ \ref{appendixA.sec} Figure\ \ref{fig:obs_qloga_allQG} and marginalized over $\log(a)$ is shown with the solid curves.  Thin dashed curves closely surrounding the solid curves illustrate the confidence intervals around the best-fit models.  The sample size of the ensemble as well as the 50th percentiles of the best-fit intrinsic ellipticity and triaxiality distributions are printed in each panel.  

Quantifying the distribution of intrinsic shapes by breaking them down into the three discrete shape families defined in Figure\ \ref{fig:shape}, we observe a significant upturn in the contribution from spheroidal shapes at the highest masses (Figure\ \ref{fig:obs_shape_allQG}, bottom panels).  Over all redshifts, the fraction of QGs with $\log(M_*)>10$ that feature elongated shapes remains small, with $\lesssim 20\%$.  The disk fraction among QGs is found to be high at intermediate ($\log (M_*) \sim 10 - 10.5$) masses at all redshifts considered, with more disparate results seen at the low-mass end, where our analysis may also be prone to incompleteness effects (in the case of oblate axisymmetric shapes, face-on objects that are round in projection will have a fainter surface brightness than their edge-on counterparts, thus at low mass and especially high redshift they may be the first to drop out of the sample).

The findings presented in Figure\ \ref{fig:obs_shape_allQG} extend the results presented by \citet{van2009}, for $0.04 < z < 0.08$ QGs from the Sloan Digital Sky Survey (DR6) out to higher redshifts ($z \sim 0.9$, corresponding to a lookback time of 7.3 Gyr).  A consistent conclusion on spheroidal shapes of the most massive QGs was also formulated by \citet{Holden2012}, who exploited a sample of QGs identified in the smaller area GEMS and COSMOS fields.  We here place these results on a statistically more robust footing, and confirm they remain intact also when considering the two-dimensional $q$ - $\log(a)$ distribution.  Rather than purely internal stellar processes to grow a massive bulge from pre-existing disk material, massive galaxies are more likely to transform into spheroidal shapes by the scrambling of orbits and violent relaxation induced via (major) mergers which destroy the disk structure of intermediate mass galaxies.

We also note that these findings are closely related to results on dynamical properties of nearby early-type galaxies as revealed by integral-field spectroscopic surveys such as SAURON \citep{Emsellem2007}, ATLAS$^{\rm 3D}$ \citep{Emsellem2011}, MaNGA \citep{Bernardi2019}, and SAMI \citep{vandeSande2021}.  These studies find so-called slow rotators, characterized by low ratios of rotational velocity compared to velocity dispersion, to be most dominant among the most massive galaxies (which are likely the product of dry mergers), and to reside in dense environments (with an associated richer merger history).  Based on simulations in a cosmological context, previous analyses concluded that, whilst it is not impossible to create such massive slow rotators without mergers, they are on average found to have experienced the highest frequency of major mergers (\citealt{Penoyre2017}; see also \citealt{Naab2014}).

\begin{figure*}
\centering
\includegraphics[width=\linewidth]{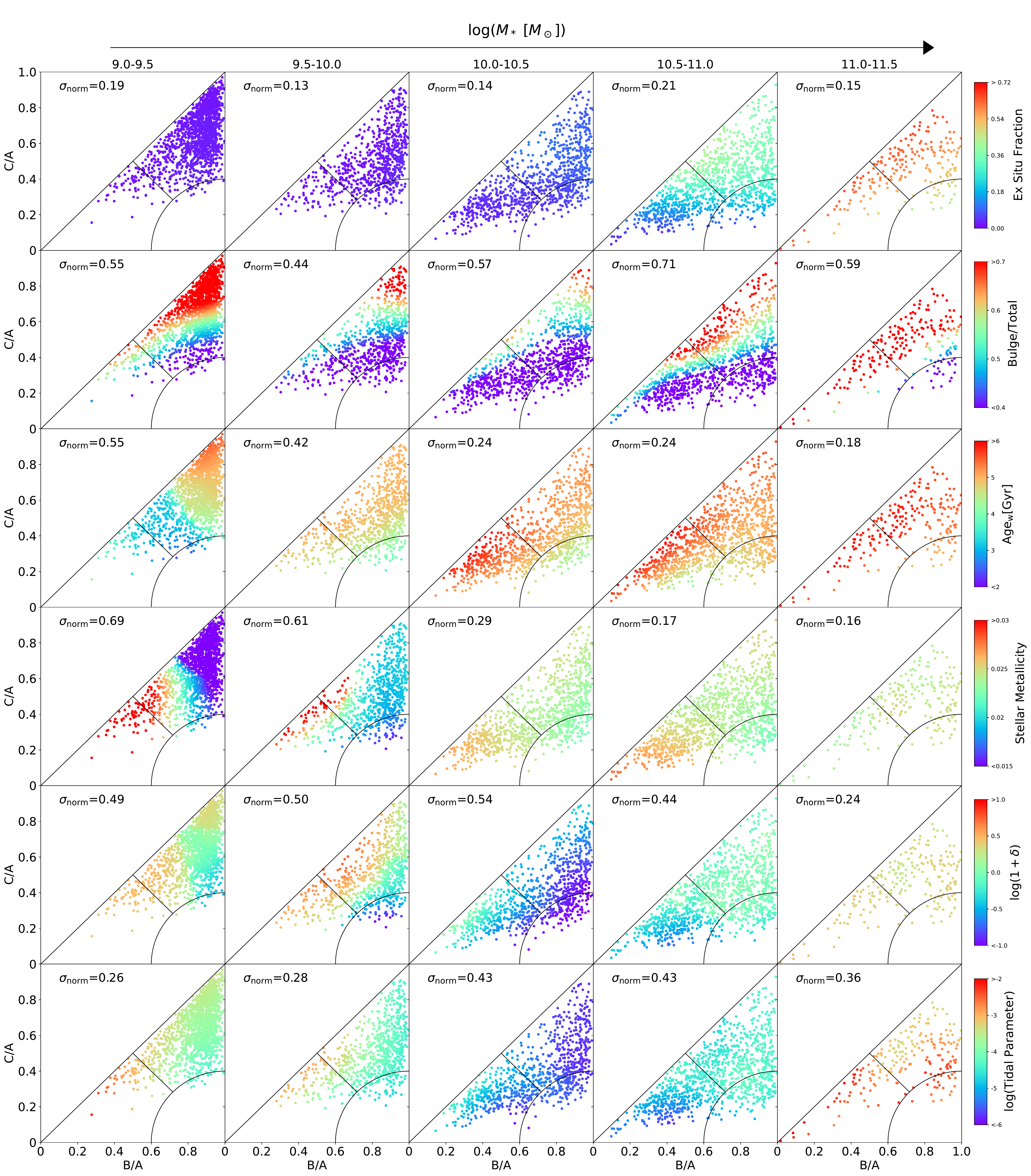}
\caption{
Intrinsic shape plane for quiescent TNG galaxies at $z=0.44$, sorted by increasing mass ({\it left to right}) and colour-coded by a series of physical properties related to their assembly history, internal structure, stellar population content, and environment.  Specifically, from top to bottom the colour coding represents the fraction of stellar mass formed ex situ, bulge-to-total ratio, the mass-weighted stellar age, stellar metallicity, overdensity, and tidal parameter.  
The colour coding applied encodes the underlying trend as revealed by a two-dimensional local regression (LOESS; \citealt{Cappellari2013}).  The parameter $\sigma_{\rm{norm}}=  \frac{\sigma(\rm{Property}-\rm{Property_{LOESS}})}{\rm{Max_{colorbar}-Min_{colorbar}}}$ quantifies the galaxy-to-galaxy scatter around the LOESS-smoothed relation, normalised by the dynamic range of the respective colourbar.  Black lines denote the same shape family regions (disky, spheroidal, elongated) as introduced in Figure\ \ref{fig:shape}.  The $z=0.44$ snapshot corresponds to the middle of the $0<z<0.9$ redshift range considered observationally.  Similar trends are recovered over the full $0<z<0.9$ range.
}
\label{fig:TNG_shape}
\end{figure*}

Turning to the TNG simulations, a similar upturn of the fraction of spheroidal galaxies among the quiescent population is seen going from intermediate masses ($\log(M_*) \sim 10.5$) to the massive end ($\log(M_*)>11$).  This is illustrated in Figure\ \ref{fig:sim_shape_allQG}, where the breakdown in shape families is based on the 3D distribution of stellar particles following the methodology outlined in Section\ \ref{shape_sim.sec}, and the uncertainties depicted by polygons are derived via bootstrapping.  We note, however, that in two other aspects the shape results obtained from TNG show marked differences from the observational results.  First, whereas a larger abundance of flat/thin shapes (i.e., with a smaller minor-to-major axis ratio $C/A$) are found among simulated galaxies of intermediate mass ($\log(M_*) \sim 10-11$), in detail they are most commonly accompanied by relatively low middle-to-major axis ratios too, placing the respective objects in the shape family of elongated sytems.  Such prolate objects make up $\sim 40\%$ of TNG QGs in this mass regime, with disky systems accounting for a more modest fraction (up to 17\%).  Second, considering TNG QGs of lower mass ($\log(M_*) \lesssim 10$), the proportion of spheroidal intrinsic shapes increases sensitively again, in contrast to the relatively flat trend observed in the real Universe (bottom panels in Figure\ \ref{fig:obs_shape_allQG}).  We confirm that both discrepancies are present in the higher resolution TNG50 simulations as well, although with smaller number statistics.  The difference between stellar 3D shapes of simulated and observed galaxies is not unique to TNG, or to the quiescent galaxy family, either.  Analyzing the output of the independent cosmological hydrodynamical simulation EAGLE, \citet{deGraaff2022} report on median projected axial ratios of simulated galaxies being relatively constant with mass, unlike what is observed.  They further find a lack of simulated counterparts to the flattened structures observed among both SFGs and QGs in the real Universe.  They tentatively attribute this difference to a pressure floor imposed in the simulation and/or numerical heating by 2-body interactions due to the limited mass of dark matter particles (see also \citealt{Ludlow2021}).  In a similar vein, \citet{Pillepich2019} illustrate how stellar structures are markedly thicker than gas disks among star-forming galaxies in TNG.

Stepping away from the classification into three discrete shape families, Figure\ \ref{fig:TNG_shape} shows the distribution of individual TNG QGs in intrinsic axial ratio space (minor-to-major $C/A$ vs. middle-to-major $B/A$).  In order to shed light on the assembly history of QGs of different mass in the simulation, each row colour-codes the objects by a different physical property.  The top row of Figure\ \ref{fig:TNG_shape} shows the fraction of each galaxy's stellar content that was formed ex-situ (i.e., not via star formation in the main progenitor branch of the merger tree).  The ex-situ fraction within simulated QGs starts to increase from $\log (M_*) \sim 10.5$, and especially rapidly above $\log (M_*) > 11$, as also reported on the basis of cosmological simulations by, e.g., \citet{RodriguezGomez2017} and \citet{Remus2021}.  Figure\ \ref{fig:TNG_shape} further shows that within a given high-mass bin the intrinsic shape and ex-situ mass fraction are correlated such that round systems tend to be associated with the highest ex-situ mass fractions, consistent with our earlier interpretation.\footnote{In detail, slight differences in the trend are notable between the $\log (M_*) \sim 10.5 - 11$ and $\log (M_*) \sim 11 - 11.5$ bin, such that the ex-situ fraction traces more closely $C/A$ in the former and triaxiality in the latter ($T = 1$ corresponding to the diagonal $C/A = B/A$).}

Since the ex-situ fraction is not a direct observable, we consider in the second row of Figure\ \ref{fig:TNG_shape} also the stellar bulge mass fraction, or bulge-to-total ratio ($B/T$).  At every mass, intrinsic shapes are clearly correlated with $B/T$, such that rounder (higher $C/A$) and more triaxial (closer to $C/A = B/A$) systems tend to be more bulge-dominated.  At high masses, this mimics the trends observed in ex-situ fractions, reflecting that the mergers bringing in external mass are also responsible for the shape transformation and bulge growth.  At lower masses, a similar trend of $B/T$ with intrinsic shape persists whereas ex-situ fractions remain low throughout shape space, suggestive of other bulge formation channels than mergers being at play.  In Appendix\ \ref{appendixB.sec}, we test and confirm that also in the observations, if we only consider the subset of observed QGs with the highest $B/T$, we recover axial ratio distributions that are consistent with rounder intrinsic shapes.  A qualitatively similar relation between bulge prominence and intrinsic shape as displayed in Figure\ \ref{fig:TNG_shape} thus appears to be present among observed QGs as well.

We further note that for TNG QGs, the bulge-dominated ($B/T > 0.5$) fraction reaches a minimum (of 23-29\% depending on redshift) in the $10<\log(M_*)<10.5$ mass bin, with the bulge-dominated fraction increasing on either side.  In contrast, the observational sample shows a bulge-dominated fraction that rises monotonically with mass (from 44-66\% to 94-99\% depending on redshift).  This qualitatively different behaviour is reminiscent of the different mass dependence of spheroidal fractions (in terms of 3D shape) presented in Figures\ \ref{fig:obs_shape_allQG} and\ \ref{fig:sim_shape_allQG}.  For full disclosure, we here caution that the comparison is not like for like.  The $B/T$ values displayed in Figure\ \ref{fig:TNG_shape} are taken from the TNG stellar circularity table\footnote{Specifically we adopt as $B/T$ for simulated galaxies the {\tt CircTwiceBelow0Frac} parameter, which is defined in detail in \url{https://www.tng-project.org/data/docs/specifications/$\#$sec5c}.} \citep{Genel2015} and were determined dynamically rather than on the basis of the 2D surface brightness distribution.  A full mock analysis of TNG galaxy images, which ideally would incorporate also effects due to any $M/L$ ratio variations, finite depth and resolution, is beyond the scope of this paper.

Finally, besides trends with ex-situ and bulge fractions, Figure\ \ref{fig:TNG_shape} (rows 3 and 4) suggests that a relation between intrinsic shapes and formation histories should also manifest itself in terms of stellar population ages and chemical enrichment levels that may vary systematically with intrinsic shape.  Massively multiplexed spectroscopic surveys may enable constraining such stellar population properties, allowing to bin by them and analyse their $q$ - $\log(a)$ distributions to test the TNG predictions in this regime.

\subsection{Dependence on environmental parameters}
\label{environment.sec}

We now turn to the role of environment in shaping galaxies, evaluating first the predictions made by the IllustrisTNG simulations, and addressing next what trends are recovered in our observational analysis.

\subsubsection{TNG predictions}
\label{TNG.sec}

The bottom two rows of Figure\ \ref{fig:TNG_shape} capture with their colour coding the variation in environmental overdensity and tidal parameter, respectively, across the $C/A$ vs. $B/A$ shape plane in bins of increasing mass.  Overall, it is apparent that it is at intermediate masses that QGs in the lowest density environments are found.  This is the same mass regime where the TNG QG distribution extends to lower $C/A$, and disk fractions are relatively elevated.  Conversely, QGs at $\log (M_*) < 10$ or $\log (M_*) > 11$ tend to be found in overdense environments.  At a given mass, dependencies on intrinsic shape are also notable, exhibiting similar patterns for overdensity and tidal parameter.  Among low-mass QGs ($\log (M_*) < 10$) for example, those with more disky shapes are found in lower density environments, whereas the environmental impact of neighbours appears to promote disk destruction leaving more triaxial or elongated shapes.  Oddly enough, this trend does not persist at $\log (M_*) \sim 10.5 - 11$, where it is in fact the elongated population (also of the lowest ex-situ fraction, relatively bulge-poor and metal-enriched) that is found to live in underdense environments.

\subsubsection{Inference from photometric surveys}
\label{env_obs.sec}

\begin{figure}
    \centering
    \includegraphics[width=\linewidth]{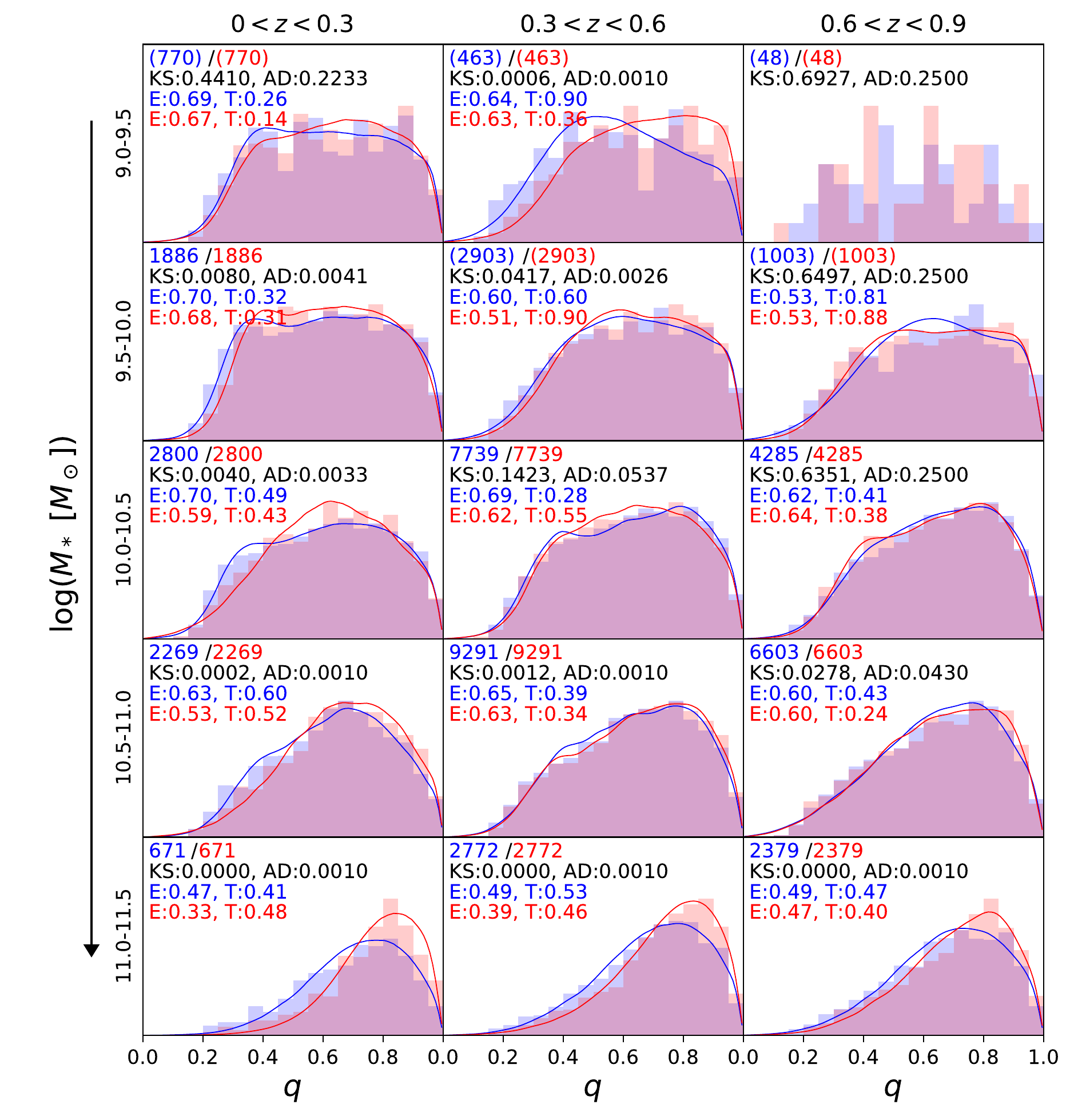}
    \includegraphics[width=\linewidth]{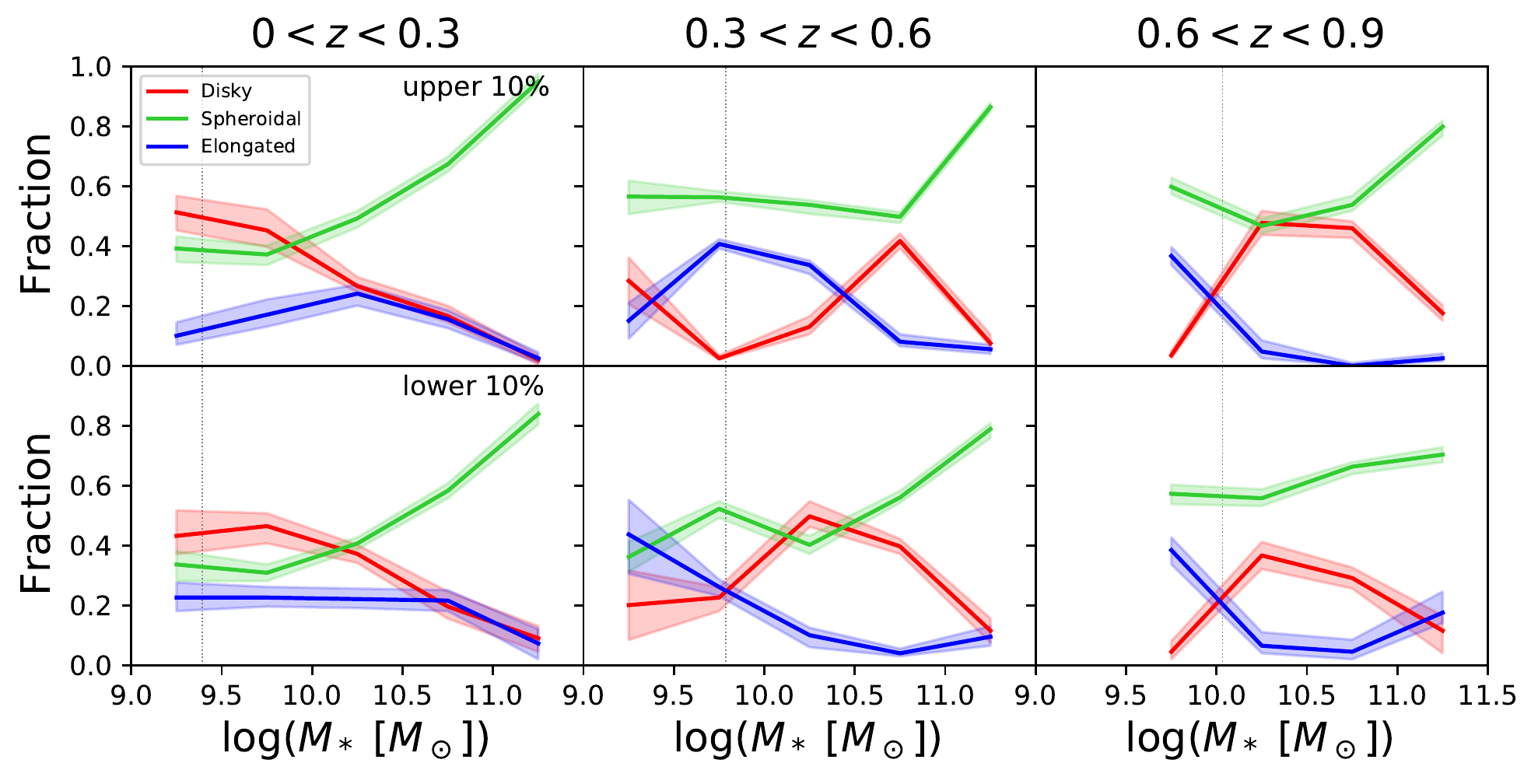}
    \caption{$q$ distribution of QGs that live in the bottom 10\% (blue) and the top 10\% (red) densest environment, split by mass and redshift. If the Kolmogorov-Smirnov (KS) or Anderson-Darling (AD) test p-values are high, then we cannot reject the null hypothesis that the two distributions are identical.  Median values of the inferred intrinsic ellipticity and triaxiality distributions are printed in each panel, as are the number of QGs comprised in each ensemble.  At the highest masses, QGs living in overdense environments are found to be rounder.}
    \label{fig:QGshape_overdensity_up_low}
\end{figure}

Turning to the observations, we recognise that even for high-quality photometric redshifts any measure of local overdensity will necessarily be diluted by foreground and/or background interlopers.  That is, the depth of the cylinder within which we quantify the local number density, deliberately chosen to scale with the characteristic $z_{\rm phot}$ uncertainty, far exceeds its dimension on the sky such as not to miss too many neighbours that have a genuine physical association.  As a consequence though, this unavoidably comes at the expense of significant contamination.  We therefore search for the presence of any environmental impact on galaxy shape by contrasting the high and low 10th percentile tails of the overdensity distribution (rather than, say, contrasting the upper versus lower half) as we can then be more confident that the two populations probe genuinely distinct environments.

Figure\ \ref{fig:QGshape_overdensity_up_low} presents these results for separate bins of mass and redshift, in a similar manner as previously done for the overall QG population (Figure\ \ref{fig:obs_shape_allQG}) and its bulge-poor subset (Figure\ \ref{fig:bulgelessQG}).  Again, while only the one-dimensional $q$ distribution is displayed for clarity, the 50th percentile of the inferred intrinsic ellipticity and triaxiality distribution quoted in each panel stems from modelling the two-dimensional $q$ - $\log(a)$ distribution, akin to what is shown in Appendix\ \ref{appendixA.sec}.

To ensure that we are comparing the intrinsic shapes for different environments at the same mass, we apply a fine mass-matching scheme.  That is, we split each mass bin into fine mass slices of 0.05 dex width, and for each such slice select the top and bottom 10\% of objects in terms of their environmental overdensity.  In doing so, we verified that the bottom 10\% always had $\log(1+\delta) < 0$.  This scheme guarantees the mass distributions of the subsamples compared are closely matched, even in the presence of a differently shaped galaxy stellar mass function between low- and high-density regimes \citep[see, e.g.,]{Tomczak2017}.

Whereas the bottom panel of Figure\ \ref{fig:obs_fquench_q}, showing the $q$ distribution as a function of local overdensity marginalizing over all masses and redshifts, showed no clear evidence for an environmental impact on galaxy shape, a different picture emerges when we consider just the most massive galaxies ($\log M_* > 11$) in Figure\ \ref{fig:QGshape_overdensity_up_low}.  Both Kolmogorov-Smirnov and Anderson-Darling tests confirm the projected shape distributions of the most massive QGs to be significantly distinct between the lowest and highest density environments, with the latter hosting the roundest objects.  At all redshifts, the inferred typical ellipticity of QGs in the densest environments are lower by $\Delta E \sim 0.1$.  This comes on top of the fact that the quenched fraction among the overall galaxy population is already enhanced in this overdense regime (Figure\ \ref{fig:overdensity}).  The dependence of galaxy intrinsic shape on environment is thus yet more pronounced if considering galaxies of all types (SFGs/QGs) jointly, as often done in studies of the morphology - density relation.

At lower masses ($\log (M_*) < 11$), any environmental dependence among QGs is far reduced, but if present in any form again corresponds to a slightly reduced number of the flattest (lowest $q$) objects in projection within overdense environments.  We conclude that environments rich in neighbours are not conducive to the preservation of thin structures.  This effect manifests itself most clearly at the high mass end, and is even present in an environmental study that is largely based on photometric redshifts.

We note that the trend of rounder massive QGs in overdense environments is present at the same level of significance when not applying the aforementioned fine mass-matching scheme, but simply contrasting the bottom and top 10th percentile in overdensity among all QGs in the $\log(M_*) = 11 - 11.5$ interval instead.  

As a further sanity check, we verified that the trend remains present when attempting to account for measurement errors on stellar mass.  These can lead to Eddington bias, and more severely so when the galaxy stellar mass function drops off more steeply, as is the case for underdense compared to overdense environments.  To do so, we parameterised a relation between the observed and sought-for intrinsic stellar mass (in log units) as a third order polynomial, and considered via fitting which polynomial coefficients would reproduce the observed mass distribution when convolving the corresponding intrinsic mass distribution with a Gaussian error kernel.  We considered $1\sigma$ errors on stellar mass of 0.1, 0.2 and 0.3 dex, and in each case derived a separate $\log(M_{\rm observed}) - \log(M_{\rm intrinsic})$ relation for QGs in underdense/overdense environments based on the mass distributions of the 10\% lowest/highest overdensity objects selected from the entire QG sample.  Applying the two relations to all QGs in underdense/overdense environments, we then repeated our fine mass-matching routine now using the inferred intrinsic masses (where the inferred intrinsic-to-observed mass ratios for massive QGs in underdense environments were slightly lower than those in overdense environments).  Again, even when accounting for Eddington bias effects due to an assumed random error on stellar mass as large as 0.2 dex, we recovered the same trend of rounder intrinsic shapes of QGs in high overdensities compared to mass-matched counterparts at low overdensity.  Only when assuming a random uncertainty of 0.3 dex does the Eddington bias become sufficiently severe that, paired with the mass dependence of intrinsic shapes (see Section\ \ref{high_mass.sec}), environmental differences at fixed intrinsic stellar mass are no longer statistically significant.  For reference, characteristic random errors on stellar mass quoted in the literature are usually at the level of $\sim 0.1 - 0.2$ dex \citep{Wuyts2009, Mitchell2013, Mobasher2015, Roediger2015, Lower2020, McLeod2021}.

Whereas round intrinsic shapes may not necessarily be equated one-to-one to the dynamically defined class of slow rotators \citep[e.g.,]{Emsellem2011}, it is interesting to note that for the most massive quiescent galaxies in the LEGA-C survey ($0.6<z<1.0$) the slow-rotator fraction depends strongly on environment, increasing from $\sim 20\%$ to $\sim 90\%$ with increasing overdensity \citep{Cole2020}.  In the local Universe, once controlling for stellar mass, any dependence of slow-rotator fraction on environment becomes more marginal to insignificant (depending on the specific environmental metric adopted; \citealt{Veale2017}).

Splitting our observed samples by the estimated tidal parameter revealed a much stronger contrast in projected shape distributions than those compared in Figure\ \ref{fig:QGshape_overdensity_up_low}.  At face value, $q$ distributions for QGs with low tidal parameters appear to be more skewed to rounder projected shapes, at all masses and redshifts.  For reasons detailed below, we do not believe these differences to reflect genuine distinctions in intrinsic shape.  We therefore refrain from showing the figure, not to mislead the reader.  We do however devote this paragraph to discussing the effect giving rise to this apparent difference, as a cautionary note on the importance of the underlying ansatz of deriving intrinsic shapes from statistical axial ratio distributions, namely that of the assumed random viewing angles.  Entering in Eq\ \ref{eq:Q_pi} for the tidal parameter is the size of the target galaxy, coming in with a steep power of 3.  For an oblate axisymmetric system (with two identical long axes and one shorter one), the semi-major axis length measured in projection matches that of the 3D structure.  However, this is not generally the case and selecting on a metric (here the tidal parameter $Q_p$; see Eq.\ \ref{eq:Q_p}) which includes the inclination-dependent projected semi-major axis length will render the selected ensemble no longer independent of viewing angle.  

To test for this effect, we considered the scenario where the intrinsic shapes of all QGs, irrespective of estimated tidal parameter, are drawn from the same intrinsic shape distribution derived in Section\ \ref{high_mass.sec}.  Assigning viewing angles randomly to our mock galaxies, calculating their tidal parameters based on the projected major-axis length and draws from the neighbour distances and masses measured for our actual QG sample, and finally splitting by low/high estimated tidal parameter, we recover a qualitatively similar distinction between their projected axial ratio distributions.  This exercise confirms our inability to address any differences in intrinsic shapes as a function of tidal parameter, as projection effects do not allow us to quantify the latter and select by it without violating the assumption of random viewing angles.  As a further check, we split our sample by a slightly modified form of the tidal parameter that excludes the projected size of the target galaxy ($A_p$ in Eq\ \ref{eq:Q_pi}), and found the difference in $q$ distributions to largely disappear.

\subsubsection{Prospects for spectroscopic surveys}
\label{env_obs_spec.sec}

The role of environment as a secondary player besides mass in dictating the star-forming/quenched state and intrinsic shape of galaxies is already evident from the crude environmental measures exploited in this study.  This prompts further research over similarly wide areas with a vastly improved redshift accuracy, to facilitate large number statistics and to probe a wide dynamic range in cosmic environments.

A new generation of vastly multiplexed fibre spectrographs will enable this desired high spectroscopic completeness over unprecedented areas, required to characterize detailed environmental metrics down to the group scale, with other than local overdensity measures a quantification of the group halo mass, central/satellite status and group-centric distance of individual objects.

4MOST \citep{deJong2019} on the European Southern Observatory's VISTA telescope will be able to simultaneously obtain spectra of up to 2400 objects distributed over a hexagonal field of view of 4 square degrees.  The instrument will also have enough optical wavelength coverage to secure velocities of extra-galactic objects over a large range in redshift, thus enabling measurements of the evolution of galaxies and their context within cosmic large-scale structure over the past half of cosmic history.

MOONS \citep{Cirasuolo2020} will be a multi-object spectrograph mounted on the Nasmyth focus of the European Very Large Telescope (VLT).  With 1000 fibres deployable over a field of view of 500 square arcmin, and a total wavelength coverage from 0.6$\mu$m to 1.8$\mu$m, it will push well-sampled environmental studies for statistically significant numbers of galaxies ($\sim 1$ million) out to cosmic noon ($1 \lesssim z \lesssim 2.5$; \citealt{Maiolino2020}).

The Subaru Prime Focus Spectrograph (PFS; \citealt{Tamura2018}) is a massively multiplexed fibre-fed optical and near-infrared three-arm spectrograph (Nfiber = 2400, $380 < \lambda < 1260$ nm, 1.3 degree diameter field of view), that will provide a powerful spectroscopic complement to the high-quality wide-field imaging by the HSC-SSP Subaru Strategic Programme exploited in this study.

With these powerful new instruments on the horizon, analyses of environmental impacts on galaxy shapes and galaxy evolution more broadly are sure to be advanced.

\section{Summary}
\label{summary.sec}

We study the intrinsic 3D shapes of quiescent galaxies over the last half of cosmic history based on their axial ratio distribution, exploiting multi-wavelength $u$-to-$K_s$ photometry from KiDS+VIKING paired with high-quality $i$-band imaging by HSC-SSP.  The sample constitutes a $>50\times$ improvement in number statistics over the largest analyses in the literature covering a similar redshift range.
Both internal galaxy properties and the environment are taken into account, including mass, redshift, photometric bulge prominence, local density and tidal parameters. For comparison, the intrinsic shapes of quenched galaxies in the TNG simulations are analyzed and contrasted to their formation history. The main findings are listed below:

\begin{itemize}
    \item Over the full $0<z<0.9$ range and in both simulations and observations spheroidal 3D shapes become more abundant at $\log(M_*)>11$, with the effect being most pronounced at lower redshifts. In TNG, these most massive quiescent galaxies feature the highest ex-situ stellar mass fractions, pointing to violent relaxation via mergers as the mechanism responsible for their 3D shape transformation.
\end{itemize}
\begin{itemize}
    \item At low masses, the quiescent galaxies produced in TNG do not feature sufficiently flattened structures.  At intermediate masses ($\log(M_*) \sim 10.5$) those simulated galaxies that have lower minor-to-major intrinsic axis ratios are more commonly found to be elongated in shape, rather than disky, unlike what is seen in observations.  Similar findings on relatively thick stellar structures of galaxies produced within cosmological simulations have been reported by \citet[][based on EAGLE]{deGraaff2022}.
\end{itemize}
\begin{itemize}
    \item In both simulations and observations, the most spheroidal intrinsic 3D shapes correspond to objects with the highest $B/T$ (determined dynamically for TNG and by 2D surface brightness profile fitting for the observations).
\end{itemize}
\begin{itemize}
    \item 
    Quiescent galaxies in denser environments tend to be more spheroidal than those in lower density environments, at least at the high mass end ($\log (M_*) > 11$).  This signature of environmental impact is observed, despite the modest accuracy of photometric redshifts (compared to spectroscopic ones) which can dilute the quantitative metrics of environment due to interloper contamination.  Further investigations of the relationship between intrinsic shapes, formation histories and environment will benefit from wide-area surveys with the next generation of vastly multiplexed fibre spectrographs (4MOST, MOONS and PFS). 
\end{itemize}
The methodology of reconstructing the 3D structure of galaxies statically from 2D images is proven to be a powerful tool. It can aid in interpreting galactic structure and its relation to a galaxy's formation and assembly history over a wide range in redshift, mass and environments, covering sample sizes that to date are hard to match with complementary dynamical measurements.

\section{Acknowledgements}
The authors thank the anonymous referee for their constructive comments that helped improve the paper.  J.Z. gratefully acknowledges support from the China Scholarship Council (grant no201904910703).  S.W. acknowledges support from the Chinese Academy of Sciences President's International Fellowship Initiative (grant no. 2022VMB0004).  The research presented here is partially supported by the National Key R\&D Program of China under grant No. 2018YFA0404501; by the National Natural Science Foundation of China under grant Nos. 12025302, 11773052, 11761131016; by the "111" Project of the Ministry of Education of China under grant No. B20019.

The KiDS production team acknowledges support from: Deutsche Forschungsgemeinschaft, ERC, NOVA and NWO-M grants; Target; the University of Padova, and the University Federico II (Naples).

The Hyper Suprime-Cam (HSC) collaboration includes the astronomical communities of Japan and Taiwan, and Princeton University. The HSC-SSP instrumentation and software were developed by the National Astronomical Observatory of Japan (NAOJ), the Kavli Institute for the Physics and Mathematics of the Universe (Kavli IPMU), the University of Tokyo, the High Energy Accelerator Research Organization (KEK), the Academia Sinica Institute for Astronomy and Astrophysics in Taiwan (ASIAA), and Princeton University. Funding was contributed by the FIRST program from Japanese Cabinet Office, the Ministry of Education, Culture, Sports, Science and Technology (MEXT), the Japan Society for the Promotion of Science (JSPS), Japan Science and Technology Agency (JST), the Toray Science Foundation, NAOJ, Kavli IPMU, KEK, ASIAA, and Princeton University. 
This paper makes use of software developed for the Large Synoptic Survey Telescope. We thank the LSST Project for making their code available as free software at  http://dm.lsst.org/.
The Pan-STARRS1 Surveys (PS1) have been made possible through contributions of the Institute for Astronomy, the University of Hawaii, the Pan-STARRS Project Office, the Max-Planck Society and its participating institutes, the Max Planck Institute for Astronomy, Heidelberg and the Max Planck Institute for Extraterrestrial Physics, Garching, The Johns Hopkins University, Durham University, the University of Edinburgh, Queen’s University Belfast, the Harvard-Smithsonian Center for Astrophysics, the Las Cumbres Observatory Global Telescope Network Incorporated, the National Central University of Taiwan, the Space Telescope Science Institute, the National Aeronautics and Space Administration under Grant No. NNX08AR22G issued through the Planetary Science Division of the NASA Science Mission Directorate, the National Science Foundation under Grant No. AST-1238877, the University of Maryland, and Eotvos Lorand University (ELTE) and the Los Alamos National Laboratory.

The IllustrisTNG simulations were undertaken with compute time awarded by the Gauss Centre for Supercomputing (GCS) under GCS Large-Scale Projects GCS-ILLU and GCS-DWAR on the GCS share of the supercomputer Hazel Hen at the High Performance Computing Center Stuttgart (HLRS), as well as on the machines of the Max Planck Computing and Data Facility (MPCDF) in Garching, Germany.

\section{DATA AVAILABILITY}
The data underlying this article were accessed from the publicly available Data Releases by KiDS+VIKING, HSC-SSP and IllustrisTNG. 
Based on observations made with ESO Telescopes at the La Silla Paranal Observatory under programme IDs 177.A-3016, 177.A-3017, 177.A-3018 and 179.A-2004, and on data products produced by the KiDS consortium. 
%
Based in part on data collected at the Subaru Telescope and retrieved from the HSC-SSP data archive system, which is operated by Subaru Telescope and Astronomy Data Center at National Astronomical Observatory of Japan.
All IllustrisTNG data can be accessed at https://www.tng-project.org/.


\bibliographystyle{mnras}
\bibliography{library} 


\appendix

\section{Modelling in the \lowercase{q - log(a)} plane}
\label{appendixA.sec}

\begin{figure*}
\centering

\includegraphics[width=0.48\linewidth]{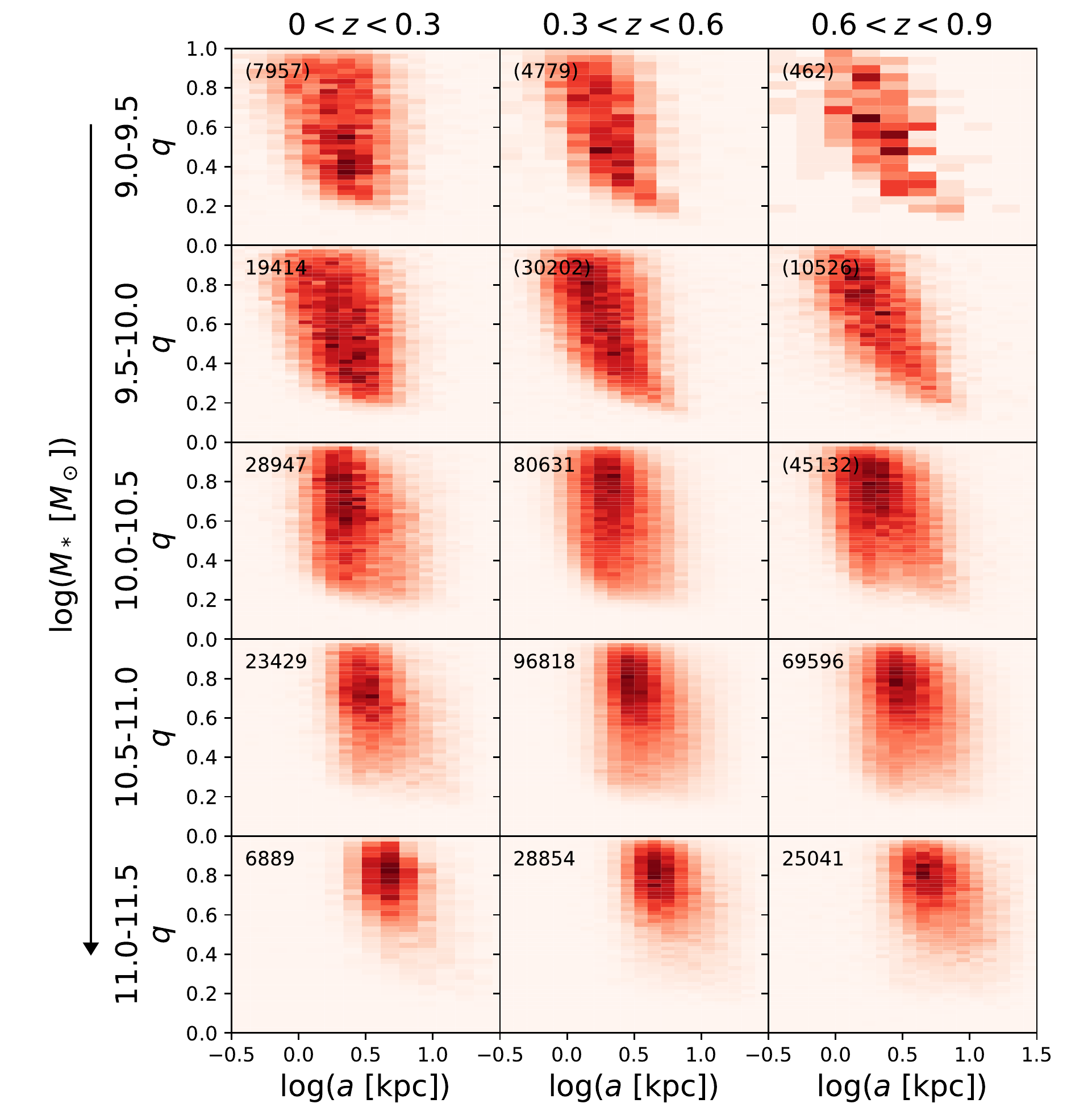}
\includegraphics[width=0.48\linewidth]{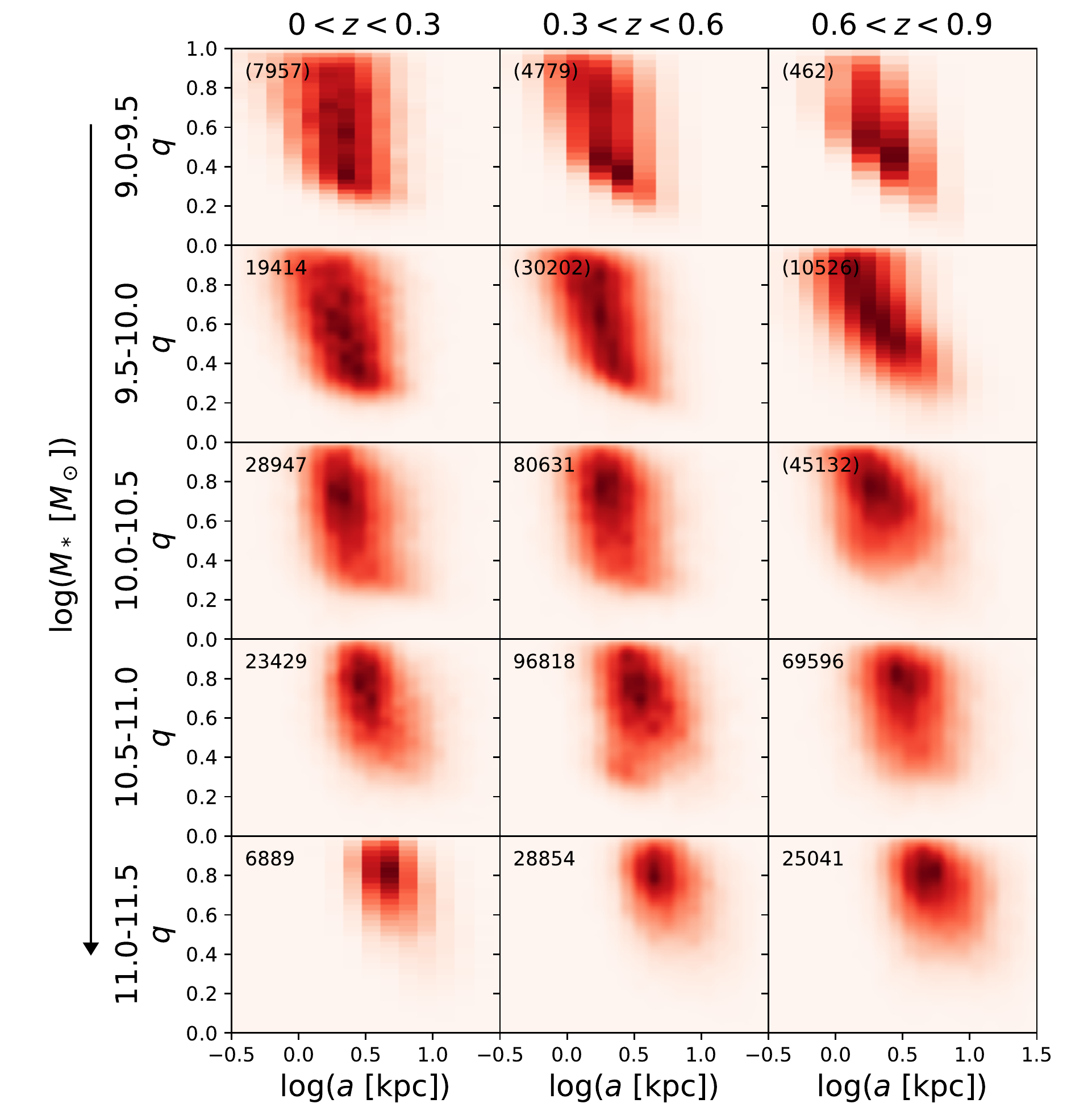}
\caption{{\it Left:} Observed 2D histograms of projected axial ratio $q$ versus projected semi-major axis length $\log(a)$ in bins of stellar mass and redshift.  {\it Right:} Best-fit model distribution derived following the steps outlined in Section\ \ref{shape_obs.sec}.  Figure\ \ref{fig:obs_shape_allQG} presents the marginalized 1D $q$ distributions for the observations and best-fit model, as well as the relative fractions of different shape families inferred from the $q - \log(a)$ modelling. 
}
\label{fig:obs_qloga_allQG}
\end{figure*}

Whereas for visualisation purposes and ease of qualitative interpretation we show plots of the projected axial ratio distributions throughout the paper, the actual quantitative modelling is in all instances carried out by minimizing residuals between the observed and modelled galaxy distribution in the two-dimensional $q$ - $\log(a)$ plane.  Figure\ \ref{fig:obs_qloga_allQG} shows these distributions for the overall QG population, binned by mass and redshift, for the observations ({\it left panel}) and best-fit model ({\it right panel}), respectively.  In the rare occasion that the marginalized $q$ distribution extracted from a best-fit model appears not to describe the observed $q$ distribution optimally, this is due to the fact that an alternative model better reproducing the $q$ distribution solely, would not represent the observed $q$ - $\log(a)$ distribution as well.

\begin{table*} 
\centering 
\resizebox{\linewidth}{!}{%
\begin{tabular}{l|l|l|l|l|l|l|l|l|l} 
\hline \hline
Redshift & $\log(M_*/M_\odot)$ & ${\langle E \rangle}^a$ & ${\langle T \rangle}^b$ & ${\langle \log(A) \rangle}^c$ & ${\sigma_E}^d$ & $\sigma_T$ &
$\sigma_{\log(A)}$ & ${\rm{Cov}}(E, \log(A))$ & $f_{\rm{population}}$\\
\hline 

$0<z<0.3$ & $9.0-9.5$ &
$0.715\pm0.065$ &
$0.105\pm0.106$ &
$0.417\pm0.010$ &
$0.066\pm0.006$ &
$0.141\pm0.062$ &
$0.202\pm0.010$ &
$0.443\pm0.109$ &
$0.665\pm0.040$
\\
&&
$0.531\pm0.186$ &
$0.155\pm0.129$ &
$0.166\pm0.025$ &
$0.187\pm0.009$ &
$0.158\pm0.045$ &
$0.183\pm0.011$ &
$1.027\pm0.101$ &
$0.335\pm0.040$ \\

$0<z<0.3$ & $9.5-10.0$ &
$0.744\pm0.054$ &
$0.230\pm0.245$ &
$0.487\pm0.007$ &
$0.054\pm0.002$ &
$0.338\pm0.042$ &
$0.201\pm0.003$ &
$0.457\pm0.031$ &
$0.520\pm0.014$
\\
&&
$0.555\pm0.184$ &
$0.147\pm0.124$ &
$0.224\pm0.006$ &
$0.187\pm0.005$ &
$0.151\pm0.025$ &
$0.172\pm0.006$ &
$0.896\pm0.057$ &
$0.480\pm0.014$ \\

$0<z<0.3$ & $10.0-10.5$ &
$0.504\pm0.232$ &
$0.550\pm0.321$ &
$0.338\pm0.003$ &
$0.244\pm0.005$ &
$0.607\pm0.051$ &
$0.161\pm0.002$ &
$0.237\pm0.022$ &
$0.509\pm0.006$
\\
&&
$0.753\pm0.056$ &
$0.439\pm0.220$ &
$0.603\pm0.004$ &
$0.055\pm0.002$ &
$0.228\pm0.014$ &
$0.218\pm0.003$ &
$0.633\pm0.028$ &
$0.491\pm0.006$ \\

$0<z<0.3$ & $10.5-11.0$ &
$0.681\pm0.095$ &
$0.454\pm0.233$ &
$0.757\pm0.002$ &
$0.094\pm0.003$ &
$0.248\pm0.032$ &
$0.228\pm0.002$ &
$0.369\pm0.022$ &
$0.504\pm0.004$
\\
&&
$0.463\pm0.194$ &
$0.370\pm0.148$ &
$0.487\pm0.002$ &
$0.199\pm0.006$ &
$0.151\pm0.016$ &
$0.148\pm0.002$ &
$0.094\pm0.024$ &
$0.496\pm0.004$ \\

$0<z<0.3$ & $11.0-11.5$ &
$0.353\pm0.153$ &
$0.640\pm0.134$ &
$0.668\pm0.007$ &
$0.155\pm0.008$ &
$0.138\pm0.041$ &
$0.133\pm0.004$ &
$0.069\pm0.054$ &
$0.580\pm0.044$
\\
&&
$0.501\pm0.209$ &
$0.786\pm0.289$ &
$0.857\pm0.018$ &
$0.214\pm0.022$ &
$0.455\pm0.126$ &
$0.171\pm0.008$ &
$0.620\pm0.095$ &
$0.420\pm0.044$ \\

$0.3<z<0.6$ & $9.0-9.5$ &
$0.717\pm0.081$ &
$0.896\pm0.318$ &
$0.428\pm0.008$ &
$0.083\pm0.006$ &
$0.777\pm0.125$ &
$0.077\pm0.009$ &
$2.481\pm0.266$ &
$0.535\pm0.026$
\\
&&
$0.466\pm0.230$ &
$0.284\pm0.158$ &
$0.219\pm0.012$ &
$0.251\pm0.017$ &
$0.172\pm0.213$ &
$0.194\pm0.008$ &
$0.694\pm0.084$ &
$0.465\pm0.026$ \\

$0.3<z<0.6$ & $9.5-10.0$ &
$0.503\pm0.214$ &
$0.263\pm0.171$ &
$0.243\pm0.003$ &
$0.222\pm0.004$ &
$0.191\pm0.015$ &
$0.192\pm0.003$ &
$0.730\pm0.036$ &
$0.589\pm0.014$
\\
&&
$0.732\pm0.074$ &
$0.788\pm0.301$ &
$0.466\pm0.004$ &
$0.075\pm0.003$ &
$0.510\pm0.036$ &
$0.124\pm0.004$ &
$1.624\pm0.064$ &
$0.411\pm0.014$ \\

$0.3<z<0.6$ & $10.0-10.5$ &
$0.735\pm0.063$ &
$0.397\pm0.206$ &
$0.522\pm0.003$ &
$0.063\pm0.002$ &
$0.217\pm0.009$ &
$0.217\pm0.002$ &
$0.494\pm0.017$ &
$0.519\pm0.008$
\\
&&
$0.486\pm0.223$ &
$0.401\pm0.228$ &
$0.295\pm0.003$ &
$0.244\pm0.005$ &
$0.249\pm0.033$ &
$0.182\pm0.002$ &
$0.153\pm0.013$ &
$0.481\pm0.008$ \\

$0.3<z<0.6$ & $10.5-11.0$ &
$0.537\pm0.184$ &
$0.410\pm0.287$ &
$0.686\pm0.003$ &
$0.188\pm0.003$ &
$0.375\pm0.013$ &
$0.207\pm0.004$ &
$0.743\pm0.020$ &
$0.655\pm0.008$
\\
&&
$0.720\pm0.066$ &
$0.112\pm0.072$ &
$0.460\pm0.003$ &
$0.065\pm0.004$ &
$0.081\pm0.009$ &
$0.147\pm0.003$ &
$0.389\pm0.029$ &
$0.345\pm0.008$ \\

$0.3<z<0.6$ & $11.0-11.5$ &
$0.531\pm0.182$ &
$0.393\pm0.207$ &
$0.888\pm0.005$ &
$0.188\pm0.007$ &
$0.221\pm0.021$ &
$0.198\pm0.003$ &
$0.576\pm0.048$ &
$0.538\pm0.009$
\\
&&
$0.391\pm0.181$ &
$0.505\pm0.072$ &
$0.665\pm0.003$ &
$0.186\pm0.005$ &
$0.073\pm0.017$ &
$0.133\pm0.002$ &
$-0.018\pm0.037$ &
$0.462\pm0.009$ \\

$0.6<z<0.9$ & $9.0-9.5$ &
$0.630\pm0.120$ &
$0.984\pm0.086$ &
$0.408\pm0.029$ &
$0.121\pm0.026$ &
$0.134\pm0.265$ &
$0.097\pm0.024$ &
$2.006\pm0.794$ &
$0.628\pm0.109$
\\
&&
$0.361\pm0.224$ &
$0.407\pm0.320$ &
$0.340\pm0.071$ &
$0.253\pm0.200$ &
$0.609\pm0.325$ &
$0.099\pm0.031$ &
$1.921\pm0.958$ &
$0.372\pm0.109$ \\

$0.6<z<0.9$ & $9.5-10.0$ &
$0.599\pm0.149$ &
$0.965\pm0.041$ &
$0.525\pm0.007$ &
$0.152\pm0.005$ &
$0.054\pm0.004$ &
$0.198\pm0.005$ &
$0.679\pm0.051$ &
$0.504\pm0.013$
\\
&&
$0.372\pm0.198$ &
$0.552\pm0.305$ &
$0.224\pm0.007$ &
$0.206\pm0.009$ &
$0.491\pm0.107$ &
$0.237\pm0.005$ &
$0.292\pm0.058$ &
$0.496\pm0.013$ \\

$0.6<z<0.9$ & $10.0-10.5$ &
$0.497\pm0.246$ &
$0.687\pm0.256$ &
$0.489\pm0.008$ &
$0.267\pm0.011$ &
$0.328\pm0.029$ &
$0.207\pm0.006$ &
$1.037\pm0.051$ &
$0.700\pm0.027$
\\
&&
$0.640\pm0.103$ &
$0.854\pm0.322$ &
$0.241\pm0.007$ &
$0.104\pm0.007$ &
$0.760\pm0.041$ &
$0.148\pm0.008$ &
$0.533\pm0.048$ &
$0.300\pm0.027$ \\

$0.6<z<0.9$ & $10.5-11.0$ &
$0.702\pm0.095$ &
$0.402\pm0.168$ &
$0.711\pm0.008$ &
$0.095\pm0.002$ &
$0.170\pm0.006$ &
$0.255\pm0.004$ &
$0.049\pm0.023$ &
$0.504\pm0.006$
\\
&&
$0.470\pm0.209$ &
$0.332\pm0.079$ &
$0.450\pm0.005$ &
$0.216\pm0.002$ &
$0.079\pm0.011$ &
$0.202\pm0.002$ &
$0.016\pm0.026$ &
$0.496\pm0.006$ \\

$0.6<z<0.9$ & $11.0-11.5$ &
$0.556\pm0.174$ &
$0.394\pm0.138$ &
$0.900\pm0.008$ &
$0.177\pm0.004$ &
$0.139\pm0.017$ &
$0.211\pm0.003$ &
$0.596\pm0.034$ &
$0.601\pm0.021$
\\
&&
$0.448\pm0.213$ &
$0.367\pm0.056$ &
$0.644\pm0.004$ &
$0.228\pm0.007$ &
$0.058\pm0.024$ &
$0.160\pm0.004$ &
$-0.036\pm0.037$ &
$0.399\pm0.021$ \\

\hline \hline  
\end{tabular}%
}
\caption{Best-fitting parameters of a two-population model for the observed quiescent galaxies in KiDS+VIKING+HSC-SSP.  For each redshift and mass bin, it is the superposition of the two populations specified in subsequent rows, weighted by $f_{\rm population}$, that best reproduces the observed $q-\log(a)$ distribution.\\
$^a E = 1 - C/A$ is the ellipticity of a galaxy.\\
$^b T = (A^2 - B^2)/(A^2 - C^2) $is the triaxiality of a galaxy.\\
$^c \langle \log (A) \rangle$ represents the average intrinsic semi-major axis length.\\
$^d \sigma_E$ is the standard deviation of the Gaussian distribution of $E$.
}
\label{tab:twopopmodel}
\end{table*}

\begin{table*} 
\centering 
\resizebox{\linewidth}{!}{%
\begin{tabular}{l|l|l|l|l|l|l|l|l|l|l|l} 
\hline \hline
Redshift & $\log(M_*/M_\odot)$ &
$E_{16}$ & $E_{50}$ & $E_{84}$ & $T_{16}$ & $T_{50}$ & $T_{84}$ &
${f_{\rm{elongated}}}^a$ & ${f_{\rm{disky}}}^a$ & ${f_{\rm{spheroidal}}}^a$ & $N_{\rm{galaxy}}$\\
\hline

$0<z<0.3$ & $9.0-9.5$ &
$0.515$ & $0.688$ & $0.771$ &
$0.057$ & $0.158$ & $0.291$ &
$0.010\pm0.013$ &
$0.724\pm0.028$ &
$0.266\pm0.020$ &
$7957$ \\

$0<z<0.3$ & $9.5-10.0$ &
$0.472$ & $0.706$ & $0.786$ &
$0.083$ & $0.231$ & $0.473$ &
$0.141\pm0.046$ &
$0.533\pm0.050$ &
$0.332\pm0.010$ &
$19414$ \\

$0<z<0.3$ & $10.0-10.5$ &
$0.394$ & $0.705$ & $0.796$ &
$0.218$ & $0.473$ & $0.751$ &
$0.176\pm0.010$ &
$0.472\pm0.020$ &
$0.357\pm0.010$ &
$28947$ \\

$0<z<0.3$ & $10.5-11.0$ &
$0.374$ & $0.614$ & $0.751$ &
$0.225$ & $0.406$ & $0.607$ &
$0.116\pm0.015$ &
$0.281\pm0.015$ &
$0.603\pm0.010$ &
$23429$ \\

$0<z<0.3$ & $11.0-11.5$ &
$0.230$ & $0.404$ & $0.608$ &
$0.432$ & $0.637$ & $0.813$ &
$0.080\pm0.015$ &
$0.045\pm0.008$ &
$0.874\pm0.013$ &
$6889$ \\

$0.3<z<0.6$ & $9.0-9.5$ &
$0.377$ & $0.661$ & $0.781$ &
$0.170$ & $0.392$ & $0.748$ &
$0.281\pm0.030$ &
$0.312\pm0.035$ &
$0.402\pm0.023$ &
$4779$ \\

$0.3<z<0.6$ & $9.5-10.0$ &
$0.371$ & $0.653$ & $0.782$ &
$0.150$ & $0.371$ & $0.713$ &
$0.216\pm0.008$ &
$0.286\pm0.018$ &
$0.497\pm0.013$ &
$30202$ \\

$0.3<z<0.6$ & $10.0-10.5$ &
$0.383$ & $0.685$ & $0.785$ &
$0.205$ & $0.415$ & $0.633$ &
$0.095\pm0.015$ &
$0.497\pm0.018$ &
$0.402\pm0.010$ &
$80631$ \\

$0.3<z<0.6$ & $10.5-11.0$ &
$0.408$ & $0.639$ & $0.764$ &
$0.087$ & $0.257$ & $0.669$ &
$0.060\pm0.003$ &
$0.462\pm0.010$ &
$0.472\pm0.010$ &
$96818$ \\

$0.3<z<0.6$ & $11.0-11.5$ &
$0.282$ & $0.468$ & $0.664$ &
$0.295$ & $0.475$ & $0.591$ &
$0.050\pm0.008$ &
$0.156\pm0.013$ &
$0.794\pm0.010$ &
$28854$ \\

$0.6<z<0.9$ & $9.0-9.5$ &
$0.337$ & $0.574$ & $0.726$ &
$0.407$ & $0.845$ & $0.956$ &
$0.352\pm0.095$ &
$0.128\pm0.088$ &
$0.503\pm0.055$ &
$462$ \\

$0.6<z<0.9$ & $9.5-10.0$ &
$0.276$ & $0.502$ & $0.691$ &
$0.362$ & $0.886$ & $0.967$ &
$0.402\pm0.015$ &
$0.010\pm0.005$ &
$0.588\pm0.015$ &
$10526$ \\

$0.6<z<0.9$ & $10.0-10.5$ &
$0.312$ & $0.568$ & $0.743$ &
$0.306$ & $0.607$ & $0.863$ &
$0.216\pm0.023$ &
$0.211\pm0.033$ &
$0.578\pm0.025$ &
$45132$ \\

$0.6<z<0.9$ & $10.5-11.0$ &
$0.378$ & $0.634$ & $0.771$ &
$0.249$ & $0.355$ & $0.492$ &
$0.065\pm0.000$ &
$0.387\pm0.010$ &
$0.548\pm0.008$ &
$69596$ \\

$0.6<z<0.9$ & $11.0-11.5$ &
$0.321$ & $0.523$ & $0.710$ &
$0.285$ & $0.377$ & $0.490$ &
$0.015\pm0.010$ &
$0.206\pm0.015$ &
$0.774\pm0.010$ &
$25041$ \\

\hline \hline  
\end{tabular}%
}
\caption{Percentiles (16th, 50th and 84th) of the ellipticity ($E$) and triaxiality ($T$) distribution of the best-fitting two-population model for observed quiescent galaxies in KiDS+VIKING+HSC-SSP, with fractional breakdown into the three shape families (see Section\ \ref{shape_obs.sec} for details on the methodology).  Each row presents the results for a bin in redshift and mass.\\
$^a$The definitions of the three shape families are depicted by the boundaries in Figure\ \ref{fig:shape}.
}
\label{tab:shapefamilies}
\end{table*}

For completeness, we provide the numerical results of the intrinsic shape modelling with two populations, as outlined in Section\ \ref{shape_obs.sec}, in two forms.  Table\ \ref{tab:twopopmodel} provides the directly fit model parameters.  Table\ \ref{tab:shapefamilies} lists population properties derived from it: percentiles of the marginalized ellipticity ($E$) and triaxiality ($T$) distribution of the best-fitting model, alongside the breakdown into shape families as visualized in Figure\ \ref{fig:obs_shape_allQG}.

\section{Rounder shapes for quiescent galaxies with de Vaucouleurs profiles}
\label{appendixB.sec}

The HSC-SSP pipeline provides an approximate decomposition of each 2D surface brightness profile into bulge and disk components, with the associated bulge-to-total ratio captured by the {\tt fracDev} parameter (here referred to as $B/T$; see \citealt{Bosch2018} for details of the measurement methodology).  Briefly, two independent fits of PSF-convolved galaxy models are carried out, one adopting an exponential profile ($n_{\rm S\acute{e}rsic} = 1$), and one with a \citet[][$n_{\rm S\acute{e}rsic}=4$]{deVaucouleurs1948} profile.  
Next, a superposition of the two best-fit profiles is fit to the galaxy image, allowing only their amplitudes to vary.  The motivation of such approach is to limit in any individual fit the number of degrees of freedom.  A comparison to single S\'{e}rsic profile fits by \citet{van2014a} in the limited area of overlap with CANDELS reveals a relatively large spread in HSC-SSP $B/T$ measurements around $n_{\rm CANDELS} \sim 2$, but a selection of high $B/T$ objects robustly retrieves objects with high ($n_{\rm CANDELS} \gtrsim 4$) single S\'{e}rsic profiles.

\begin{figure}
\centering
\includegraphics[width=\linewidth]{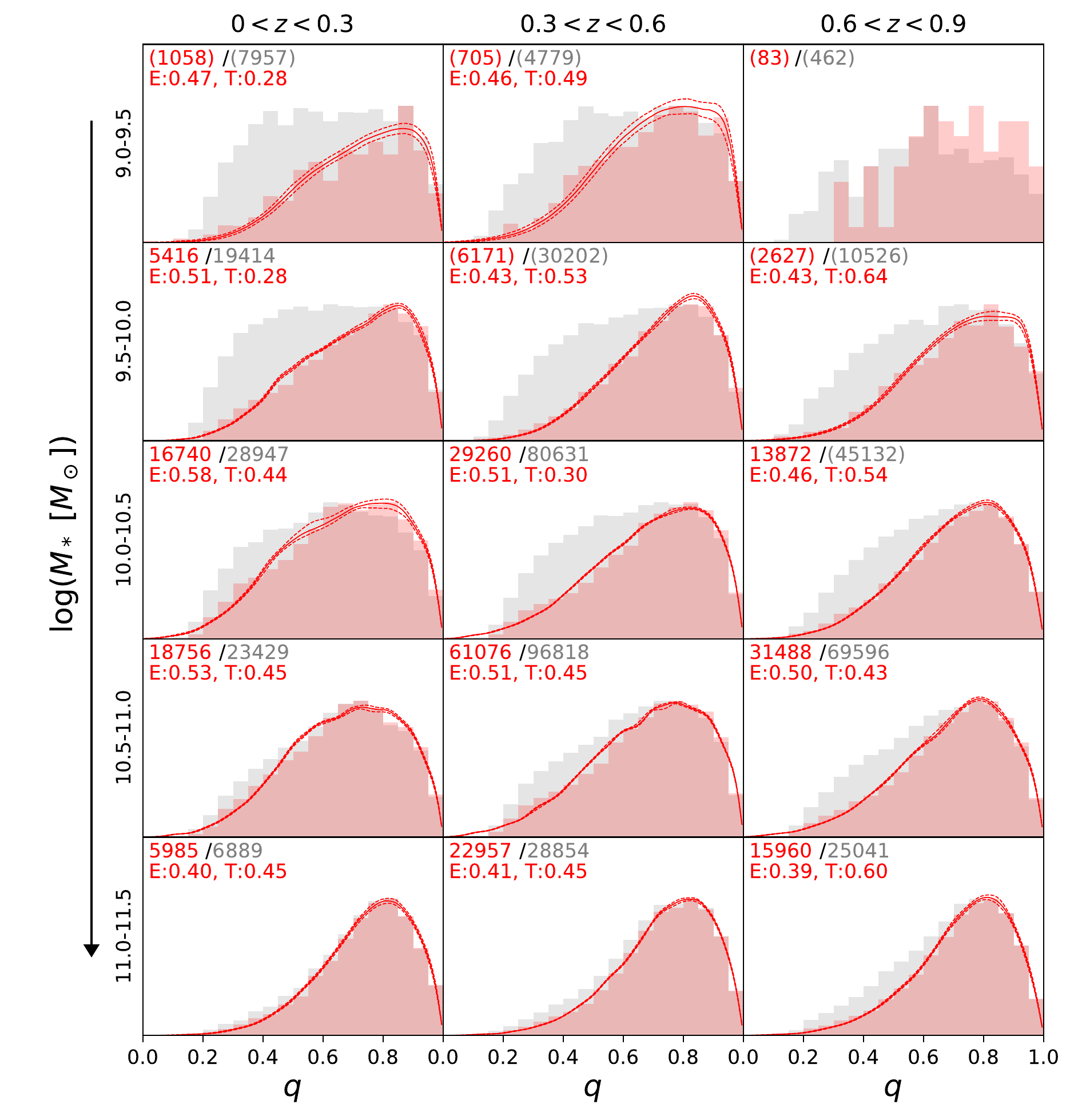}
\includegraphics[width=\linewidth]{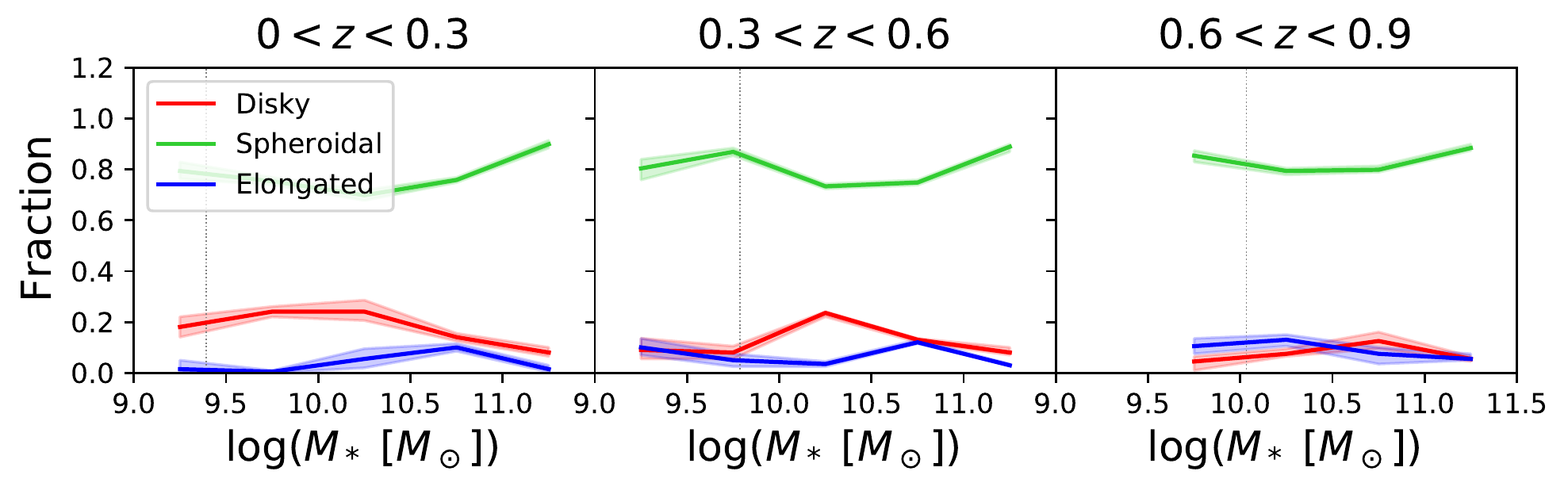}
\caption{Projected axial ratio distribution of QGs with $B/T=1$ in KiDS+VIKING+HSC-SSP (red), along with the results from their intrinsic shape modelling.  For reference, the $q$ distribution of all QGs is reproduced from Figure\ \ref{fig:obs_shape_allQG} in grey.
}
\label{fig:BT1_shape}
\end{figure}

Here, we consider the subset of QGs with $B/T = 1$, and show their axial ratio distribution in bins of mass and redshift in Figure\ \ref{fig:BT1_shape}.  Best-fit models from $q-\log(a)$ modelling are overplotted, and for reference the axial ratio distribution of all QGs is reproduced from Figure\ \ref{fig:obs_shape_allQG} in grey.  At every mass, this subset of "pure bulge" objects identified by their 2D surface brightness profiles is inferred to have rounder than average intrinsic 3D shapes.  A qualitatively similar trend is observed among simulated QGs in TNG, where the intrinsically roundest galaxies in each mass bin tend to feature above-average bulge mass fractions (second row of Figure\ \ref{fig:TNG_shape}).

\end{document}